\documentclass[aps,twocolumn,superscriptaddress, prl]{revtex4-1}

\usepackage[latin1]{inputenc}
\usepackage{amsmath}
\usepackage{amsfonts}
\usepackage{amssymb}
\usepackage{graphicx}
 \usepackage{color}        
 \usepackage{soul}
 \usepackage{verbatim}
 \usepackage{subfigure}

\newcommand{\defeq}{:=}

\newcommand{\bra}[1]{\langle #1|} 
\newcommand{\ket}[1]{\left|#1\right\rangle}

\newcommand{\abs}[1]{{\left\vert#1\right\vert}}

\newcommand{\exx}[1]{\ensuremath{\langle #1 \rangle}}

\begin{document}
\title{Quantum-enhanced capture of photons using optical ratchet states}
\author{K. D. B. Higgins}
\affiliation{Department of Materials, Oxford University, Oxford OX1 3PH, United Kingdom}
\author{B. W. Lovett}
\email{bwl4@st-andrews.ac.uk}
\affiliation{SUPA, School of Physics and Astronomy, University of St Andrews, KY16 9SS, United Kingdom}
\affiliation{Department of Materials, Oxford University, Oxford OX1 3PH, United Kingdom}
\author{E. M. Gauger}
\email{e.gauger@hw.ac.uk}
\affiliation{SUPA, Institute of Photonics and Quantum Sciences, Heriot-Watt University, EH14 4AS Edinburgh, United Kingdom}
\date{\today}
\begin{abstract}
Natural and artificial light harvesting systems often operate in a regime where the flux of photons is relatively low. Besides absorbing as many photons as possible it is therefore paramount to prevent excitons from annihilation via photon re-emission until they have undergone an irreversible energy conversion process. 
Taking inspiration from photosynthetic antenna structures, we here consider ring-like systems and introduce a class of states we call ratchets: excited states capable of absorbing but not emitting light. This allows our antennae to absorb further photons whilst retaining the excitations from those that have already been captured. Simulations for a ring of four sites reveal a peak power enhancement by up to a factor of 35 under ambient conditions owing to a combination of ratcheting and the prevention of emission through dark-state population. In the slow extraction limit the achievable power enhancement due to ratcheting alone exceeds 20\%.
\end{abstract}

\maketitle

{\it Introduction --} The absorption of light and prevention of its reemission is essential for the efficient operation of solar energy harvesting devices~\cite{kippelen09}. From the many causes of device inefficiency, few are as fundamental and seemingly insurmountable as energy loss via radiative recombination: any absorption process must have a companion emission process. This inherent absorption inefficiency is a result of the principle of `detailed balance' and constitutes a key contribution to the famous Shockley-Queisser limit \cite{shockley1961}. However, pioneering work by Scully showed that it is possible to break detailed balance, given an external source of coherence \cite{scully2010}; later work showed that this can be achieved by clever internal design alone, by using an optically dark state to prevent exciton recombination~\cite{creatore2013, zhang2015, yamada2015, fruchtman2016}. Such dark states are populated passively if the energy separation between dark and bright states falls into the vibrational spectrum of the absorbing nanostructure: dissipation then preferentially mediates transfer into states from which optical decay cannot occur. This is thought to play a role in photosynthetic light harvesters \cite{mukai1999, pearlstein1985}, e.g.~by means of dynamic localisation reducing the effective optical dipole strength~\cite{fidler2014}. 

However, time spent in dark states is `dead time' with respect to absorbing further photons.  
In organic light harvesting systems the time needed to extract an exciton from a dark state and turn it into useful energy is often orders of magnitude slower than typical light absorption rates \cite{caruso2012, blankenship02}, and this results in the loss of any subsequently arriving photons. In this Letter, we show that certain exciton states, which we will call {\it ratchets}, can enhance photocell efficiency by enabling absorption of these subsequent photons, while preventing emission.

\begin{figure}[t]
\includegraphics[width=\linewidth]{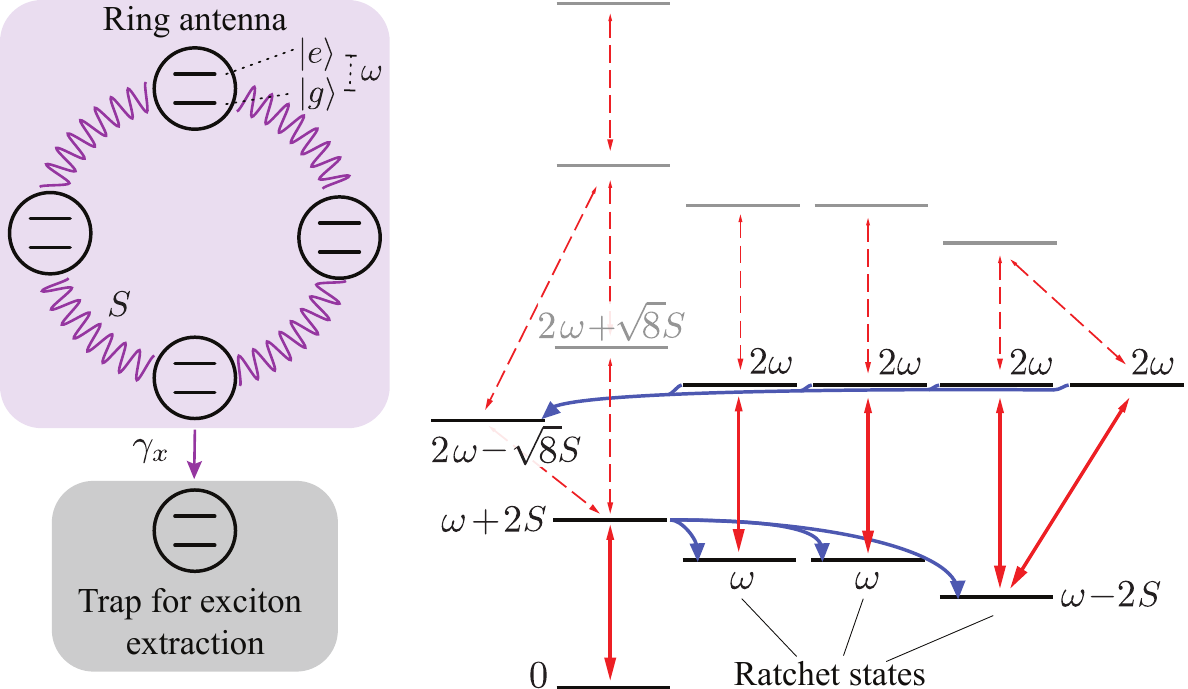}
\caption{ \label{fig1} Left: $N=4$ ring structure with attached trap. Right: Energy level diagram of this ring showing dipole-allowed optical transitions (red). Solid lines denote primary absorption transitions for the ratcheting cycle, others are indicated with thinner dashed lines. Only relevant phonon transitions are shown (blue). Rarely populated levels are shown in grey.}
\end{figure}

{\it Model --} We consider a ring of $N$ identical two-level optical emitters (see Fig.~\ref{fig1}), with nearest pairs coupled by a transition dipole-dipole interaction, governed by the Hamiltonian ($\hbar=1$):
\begin{equation}\label{eq:Ham}
H_{s} = \omega \sum_{i=1}^{N} \sigma_{i}^{+} \sigma_{i}^{-}+ S \sum_{i=1}^{N}( \sigma_{i}^{+} \sigma_{i+1}^{-} + \sigma_{i+1}^{+} \sigma_{i}^{-} ) \, , 
\end{equation}
where $\omega$ is the bare transition energy of the sites, $S$ is the hopping strength, the $\sigma^\pm_i$ denote the usual raising and lowering operators which create and destroy an excitation on site $i$, and $\sigma_{N+1}= \sigma_{1}$. For a single exciton, the eigenstates are equal superposition of excitons localized at the different ring positions, with each being characterized by a relative phase  $k_i = 2 \pi j/N$ with $ j \in 0, 1 ... N-1  $ between exciton states in adjacent positions.
Full diagonalization can be achieved by the successive application of the Jordan-Wigner and Fourier transforms \cite{jordan1928, tokihiro1993},
whence the Hamiltonian becomes $H_{s} =\sum_{K} \lambda_{K} \, \ket{K} \bra{K}$. $K$ describes the eigenstate with energy $\lambda_{K}$  and each $K$ now corresponds to a set of the phase factors $k_i$, one for each excitation [see Supplementary Information (SI)~\cite{supinf}].  
The eigenstates fall into a series of bands, where each state within a band contains the same number of excitons $n =  \abs{K}$, with $\abs{K}$ the number of elements in set $K$.

Optical transitions connect eigenstates differing by one exciton with rates proportional to $\Gamma_{K,K'} = \abs{ \bra{K'}J^{+} \ket{K}}^{2} $ where $J^{\pm} = \sum_{i=1}^{N} \sigma^{\pm}_{i} \,$. An explicit analytical expression for $\Gamma_{K,K'}$ can be derived~\cite{tokihiro1993,spano1991} and is given in~\cite{supinf}. Only the fully symmetric $k_i=0$ single exciton eigenstate has a dipole matrix element with the ground state, since here the transition dipoles interfere constructively. The other $N-1$ single exciton eigenstates do not couple to the ground state via light with a wavelength that is much longer than the separation between sites. However, these $N-1$ states do have dipole-allowed transitions to the second exciton band. They therefore have the potential for preventing re-emission of light, while still allowing further absorption of photons, creating more excitons that in a photovoltaic circuit can be converted into useful photocurrent. Such states are examples of ratchet states $\ket{K_R}$, which have the simultaneous properties:
\begin{eqnarray}\label{rat_def}
\Gamma_{K_R}^{-} \defeq \sum_{K'} \Gamma_{K_R,K'}  \, \delta_{\abs{K_R}, \abs{K'} +1} = 0 \, , \\   
\Gamma_{K_R}^{+} \defeq \sum_{K'} \Gamma_{K_R,K'}  \, \delta_{\abs{K_R}, \abs{K'}-1} > 0 \, , 
\end{eqnarray}
where $\Gamma_{K_R}^{-(+)}$ is the sum of all transition rates connecting the ratchet eigenstate $\ket{K_R}$ to the adjacent band below (above) and $\delta_{i,j}$ is the Kronecker symbol. 

For $N=4$ and $S > 0$ the bright state in the single exciton band has the highest energy $\omega + 2S$, and the three ratchet states have energies $\omega, \omega$ and $\omega - 2S$, see Fig.~\ref{fig1}b and~\cite{supinf}. 
In a molecular system, the ratchet states can be reached following excitation into the bright state when the exciton loses a small amount of energy to molecular phonons. Importantly, such vibrational modes have much shorter wavelengths than optical modes and can often be assumed to be local to each site. Unlike the photon field, the phonon interaction then breaks the symmetry and enables intraband transitions.

{\it Dynamics --} In order to model the dynamics we use the QuTiP package~\cite{johansson2013} to construct full open system dynamics in Bloch-Redfield theory, for each kind of environmental interaction. We always retain the full form of the resulting dissipator tensor since the secular approximation cannot be applied when a system has bands of closely spaced levels which may be more closely spaced than the corresponding dissipation rates~\cite{eastham16, higgins2014}. 

First, consider the dynamics due to the light-matter interaction Hamiltonian
\begin{equation}\label{eqn:lightmatter}
\hat{H}_{s-l}= \sum_{\bf Q} \sum_{i=1}^{N} G_{\bf Q} (\hat{\sigma}^+_i\hat{b}_{\bf Q}+\hat{\sigma}^-_i\hat{b}^{\dagger}_{\bf Q}) \,,
\end{equation}
where  $\hat{b}^{\dagger}_{\bf Q}$ and $\hat{b}_{\bf Q}$ are creation and annihilation operators for photons of wavevector ${\bf Q}$. The photon spectral density is constant across all transition energies, and enters into the master equation dissipator only via the single emitter decay rate $\gamma_o$. Absorption and stimulated emission terms are weighted by the appropriate Bose-Einstein factor $N(\omega_\Omega, T_o) = 1/(\exp[\omega_\Omega/ (k_B T_o)] -1)$, representing the thermal occupancy of the optical mode with interband transition frequency $\omega_{\Omega}$ and at temperature $T_o$ ($=5800$~K for solar radiation). 

Second, the exciton-phonon interaction is accounted for through~\cite{mahanbook}
\begin{equation}\label{eqn:phonon_coupling}
\hat{H}_{s-p}= \sum_{i=1}^{N}  \sum_{\bf q} g_{{\bf q}, i} \hat{\sigma}_{i}^{z}(\hat{a}_{{\bf q},i}+\hat{a}^{\dagger}_{{\bf q},i}) \,,
\end{equation}
where for each site $i$ $\hat{a}_{\bf q}$ and $\hat{a}^\dagger_{\bf q}$ are the creation and annihilation operators for phonons of wavevector ${\bf q}$ with energy $\omega_{\bf q}$ and exciton-phonon coupling $g_{\bf q}$. This interaction commutes with $\sum_i\sigma^z_i$ and so only mediates transitions within each excited band. 
For simplicity, we again assume a spectral density $J(\omega)$ that is approximately constant across all intraband transition energies $\omega_\Pi$, such that $J(\omega_\Pi)=\gamma_p$. We calculate appropriate phonon matrix elements for each transition, letting absorption and stimulated emission terms carry the appropriate Bose-Einstein factors $N(\omega_\Pi, T_p)$, with $T_p$ the ambient phonon temperature.
Reflecting the fact that phonon relaxation typically proceeds orders of magnitude faster than optical transition rates~\cite{brinks2014}, we fix $\gamma_{p}= 1000\gamma_{o}$. This separation of timescales allows population to leave the bright states before reemission occurs.

To complete the photovoltaic circuit, we include an additional `trap site' $t$, with excited state $\ket{\alpha}$ and ground state $\ket{\beta}$, where excitons are irreversibly converted into work. The trap Hamiltonian is $H_{t}=\omega_{t}\hat{\sigma}_{t}^{+}\hat{\sigma}_{t}^{-} $, where $\omega_t$ represents its transition frequency, and $\sigma_t^- = \ket{\beta}\bra{\alpha}$. Henceforth, we shall consider the joint density matrix of ring system and trap, $\rho = \rho_{s} \otimes \rho_{t}$. 
Throughout the paper we use an incoherent hopping from a single site in the ring to the trap (see SI~\cite{supinf} for an alternative model), and we fix the trap energy to be resonant with the state in which population is most likely to accumulate --- unless otherwise stated, this is at the bottom of the first exciton band, $\omega_{t} = \omega - 2S$. 

\begin{table}
\centering
\begin{tabular}{  c c c }
\hline \hline 
Parameter \,\,\, & Symbol   & Default value \\ [0.5ex] \hline 
Atomic transition frequency &$\omega$ & $1.8$ eV - 2S\\ 
{\it Hopping strength} &$S$ &0.02 eV\\ 
Spontaneous emission rate & $\gamma_{o}$ & $1.52 \times10^{9}$~Hz $\equiv 1~\mu$eV \\ 
Phonon relaxation rate & $\gamma_{p}$ & $10^{3}\gamma_{o}$\\ 
{\it Extraction rate} &$\gamma_{x}$ & $10^{-1}\gamma_{o} $ \\
{\it Photon bath temperature}  &$T_o$ & 5800 K  \\ 
{\it Phonon bath  temperature} &$T_{p}$ &  300 K \\   [0.5ex] \hline
\end{tabular}
\caption{Default model parameters; 
italic rows are plot parameters in certain Figures, as stated in the relevant captions. } 
\label{tab1} 
\end{table}
\begin{figure}[t!]
\centering
\includegraphics[width=\linewidth]{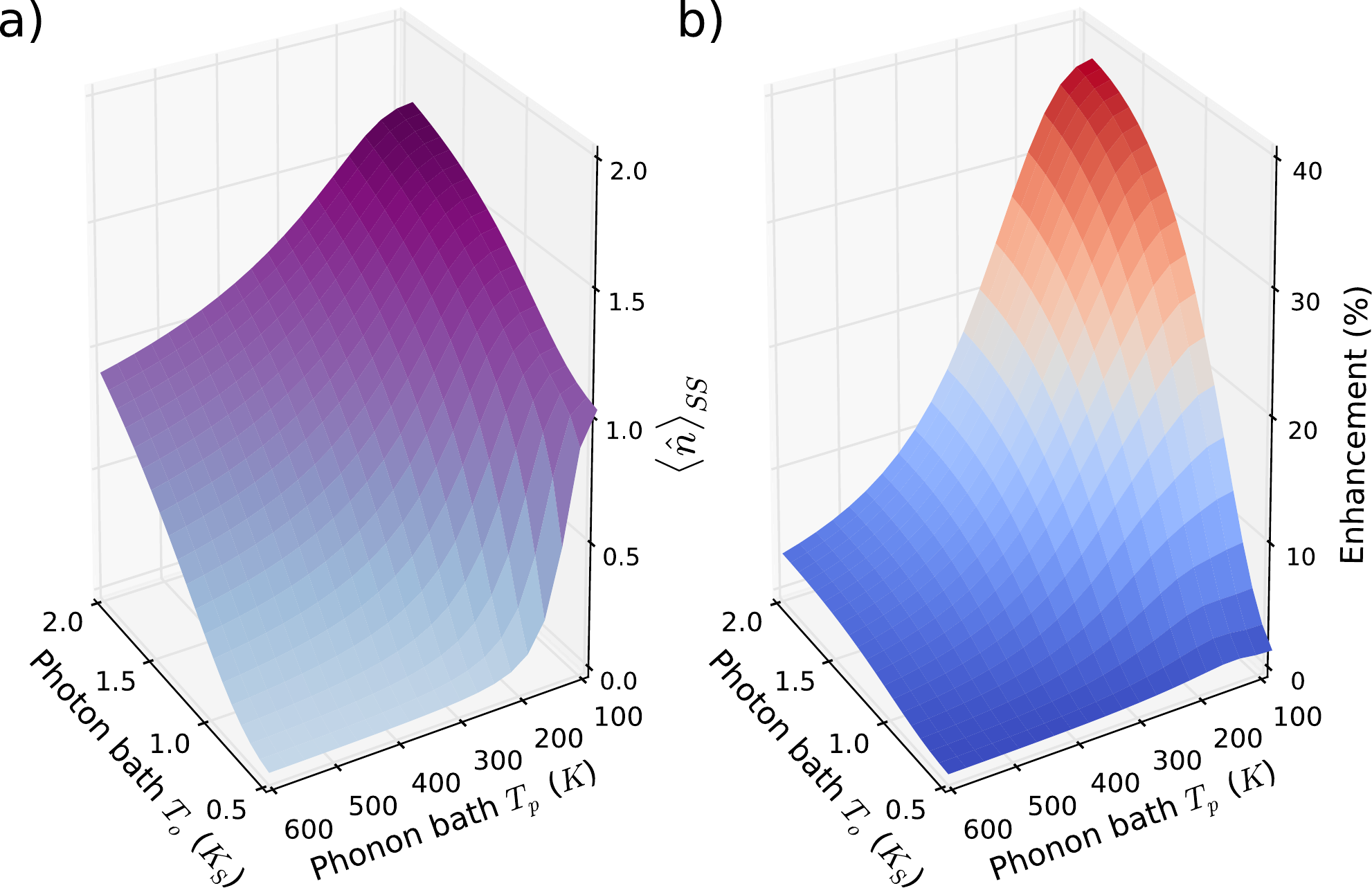}
\caption{ \label{fig2} (a) Steady-state exciton population and (b) relative ratchet enhancement compared to the FD scenario (see text), both as functions of the optical and phononic bath temperatures, and without any trap. The former is given in solar temperature units, $K_{S}=5800$ K, other parameters are as in Table \ref{tab1} except that $\gamma_{x} = 0$.}
\end{figure}

The theory of quantum heat engines~\cite{scully2011,creatore2013,dorfman2013} provides a way for us to assess the current and voltage output of the ring device: The action of a load across the device is mimicked by the trap decay rate $\gamma_t$, which varies depending on the load resistance. The current is simply $I = e \gamma_{t} \exx{\rho_{\alpha}}_{ss}$, with $\exx{\rho_{\alpha}}_{ss}$ the steady state population of the excited trap state. The potential difference seen by the load is given by the deviation of the trap's population from its thermal distribution \cite{dorfman2013,creatore2013}: 
\begin{equation}\label{eq:voltage}
eV =  \hbar\omega_t +k_{B}T_{p} \ln\left({\frac{ \exx{\rho_{\alpha}}_{ss}}{ \exx{\rho_{\beta}}_{ss}}}\right) \, , 
\end{equation}
where $k_{B}$ is Boltzmann's constant and $T_{p}$ the (ambient) phonon bath temperature. To study the relationship between current, voltage, and therefore power, we alter the trapping rate $\gamma_{t}$. Letting $\gamma_t \to 0$ or $\infty$ leads to the well-known open and short circuit limit,  respectively, however, our main interest in the following is in the region near the optimal power output point of the device.

{\it Results --} We proceed by finding the steady state of the following master equation which incorporates the optical and phononic dissipator as well as the trap decay process:
\begin{align}\label{rat_ME}
\dot{\rho} =  -i[H_{s} + H_{t}, \rho]  + D_{o}[\rho]  + D_{p}[\rho]  + D_{t}[\rho]  + D_x[\rho]\, ,
\end{align}
where $D_{o}[\rho]$, $D_{p}[\rho]$ are the Bloch Redfield dissipators for the photon and phonon fields discussed above, and 
$D_{t}[\rho] = \gamma_t(\sigma^-_t \rho  \sigma^+_t -\frac{1}{2}\{\sigma^+_t\sigma^-_t, \rho\})$ is a standard Lindblad dissipator representing trap decay. $D_x[\rho]$ represents exciton extraction via incoherent hopping at rate $\gamma_x$ from a single site on the ring to the trap (see Fig.~\ref{fig1} a).

To expose the effect of the optical ratchets we now contrast the full model of Eq.~\eqref{rat_ME} with two other artificial cases that exclude the possibility of ratchets: In the first, we remove all phonon-assisted relaxation ($\gamma_{p}=0$), meaning the system only undergoes optical transitions along the Dicke ladder of bright states \cite{higgins2014} -- we term this `no phonons' (NP). In the second construct, the ratchet states are rendered fully dark by setting all upwards transition matrix elements from them to zero, a scenario we refer to as `forced dark' (FD). All results in this section are based on the steady state of the system for $N=4$, obtained by an iterative numerical method performed with QuTiP \cite{johansson2013}, and using the default parameters from Table \ref{tab1} unless explicitly stated otherwise. Our choice of $\omega=1.8$~eV$-2S$ differs from that in previous works~\cite{creatore2013, zhang2015}: our subtraction of $2S$ ensures that the highest (bright state) level in the first band always remains at 1.8~eV and absorbs at a fixed frequency. 

In Fig.~\ref{fig2} we display the steady-state exciton population of the system without trapping as a function of optical and phonon bath temperatures. We compare the ratchet model to FD, finding that the ratchet enhancement is greatest in the hot photons, cold phonons regime. This is to be expected: hot photons quickly promote the system up the excited bands via ratchets, with cold phonons allowing a one-way protection of gained excitation energy. For FD states at low phonon temperature the exciton population plateaus near one, as most population ends up trapped in the state at the bottom of the first band (see \cite{supinf} for the FD data and an extended discussion). By contrast, the ratchet states keep absorbing, allowing the steady state population to rise toward the infinite temperature limit of $2$ (for $N=4$). 

\begin{figure}[t!]
\includegraphics[width=0.85\linewidth]{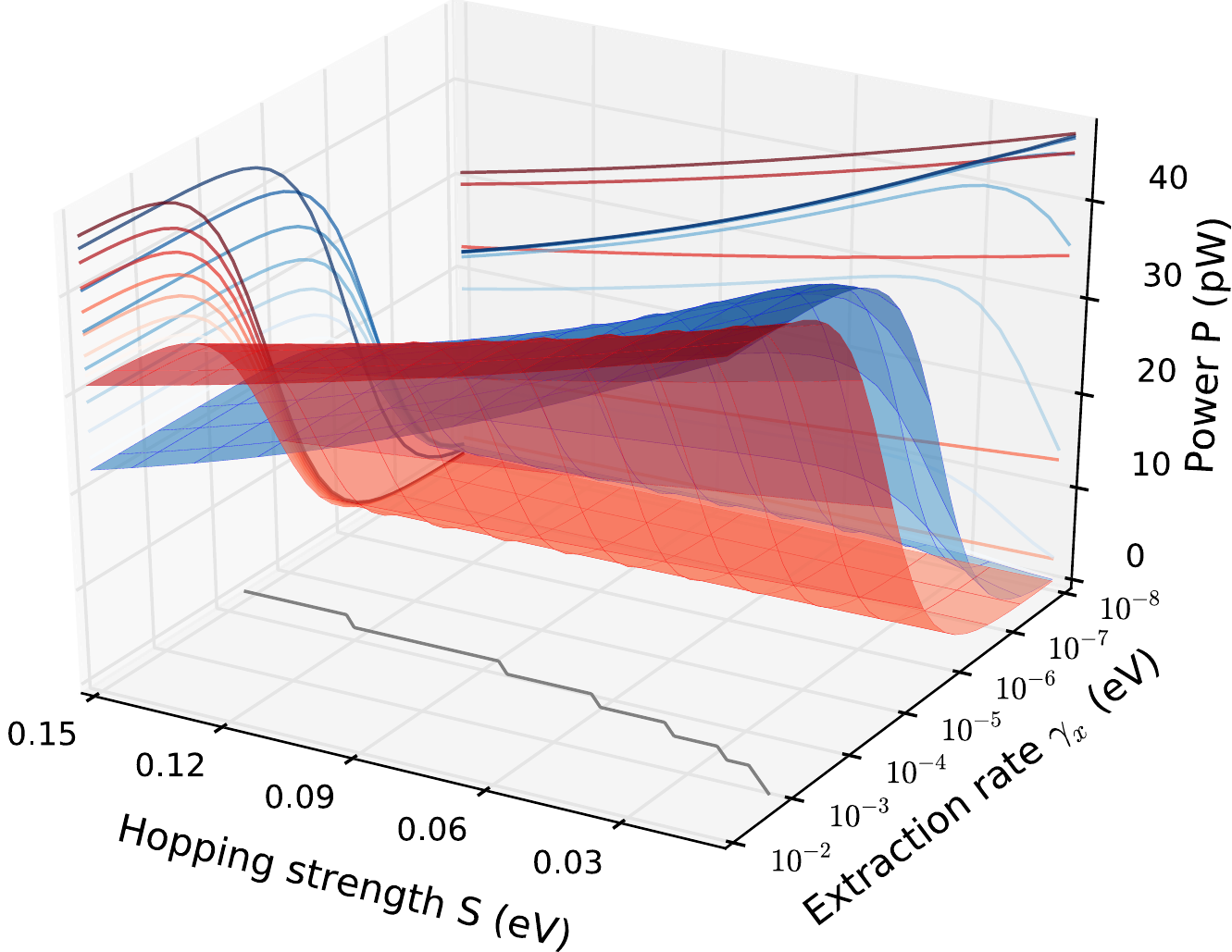}
\caption{ \label{fig3} Absolute power output in the ratcheting (blue surface) and NP (red surface) cases as a function of (single site) extraction rate $\gamma_x$ and hopping strength $S$. Other parameters are as in Table I. For each point, the optimal trap decay rate is found by numerical search. The sets of curves projected onto the side walls of the 3D plot are cuts through the data for fixed values of the appropriate parameter. The two surfaces cross in the region where the bottleneck is lifted, with ratcheting generating more power for a severe and intermediate bottleneck. The line at which the curves cross is projected onto the bottom ($xy$) plane of the figure.}
\end{figure}

Using the heat engine model, we can now explore the performance of the four site ring as an energy harvester. We will focus on the dependence of the power output on two parameters: the coupling $S$, and extraction rate $\gamma_x$. 
For each parameter set, the trap decay rate $\gamma_t$ is varied for maximum power output; example power traces as a function of $\gamma_t$ can be found the SI~\cite{supinf}.   

\begin{figure}[t!]
\includegraphics[width=\linewidth]{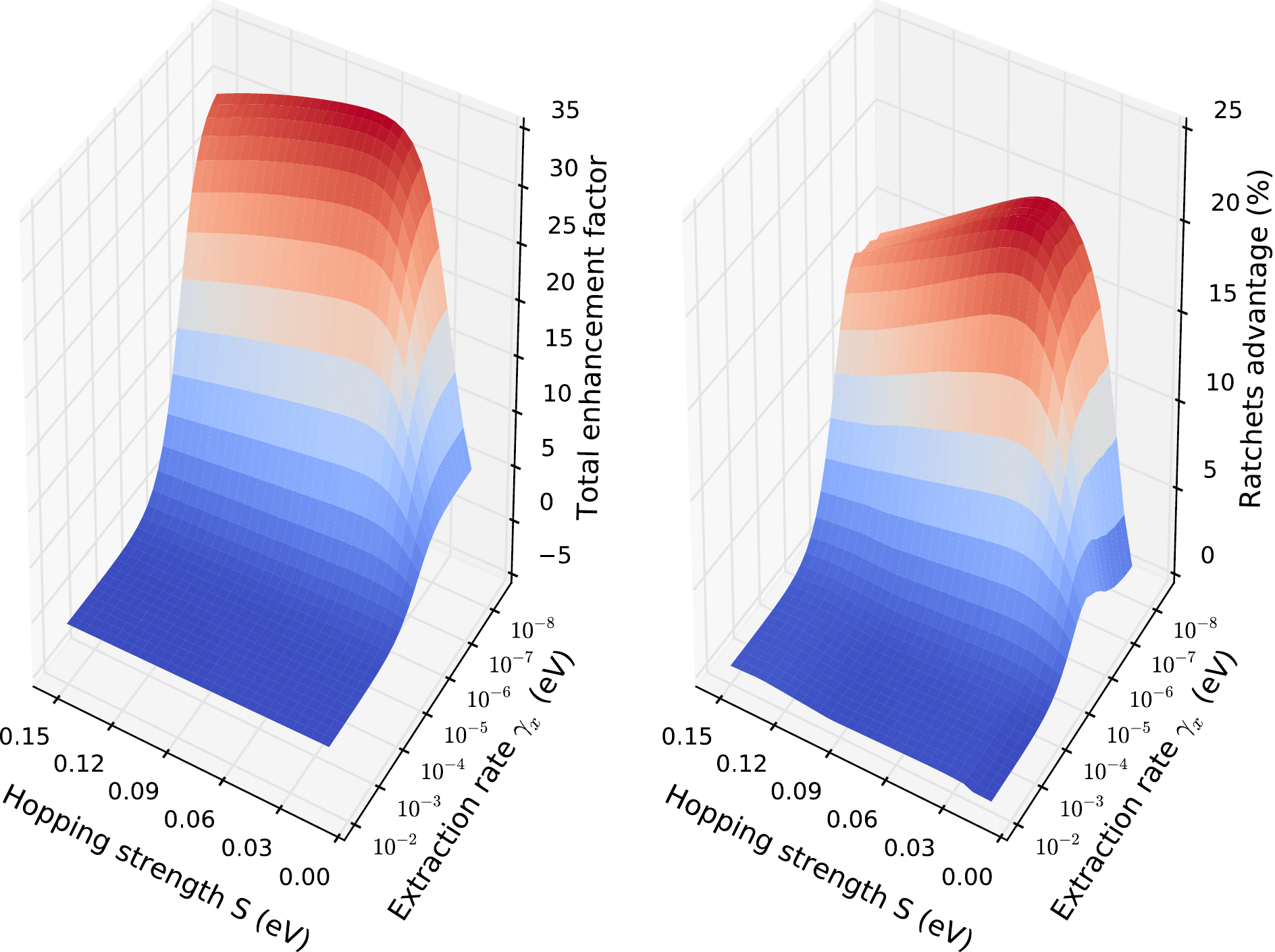}
\caption{ \label{fig4} 
Enhancement of power output for ratchet states over NP (left, displayed as a multiplicative factor) and FD (right, displayed as a percentage), as a function of both hopping strength $S$ and extraction rate $\gamma_x$. Parameters are as in Fig.~\ref{fig3}.
}
\end{figure}

In Fig.~\ref{fig3} we display the {\it absolute} value of the power generated for the ratchet and NP scenarios, whereas in Fig.~\ref{fig4} we display the {\it relative} power output of ratchet states over each of the two artificial scenarios. To ensure fair comparison, for NP we set the trap energy to be resonant with the bright state at $\omega+2S$, since the ratchet states at lower energy are inaccessible here.
All three scenarios have lower output power when the trapping (charge separation) rates are low, since this creates a bottleneck in the cycle and limits the size of the photocurrent. 

However, the ratchet states hold a great advantage over the other scenarios in this bottlenecked region, since this is precisely the situation in which excitations need to be held for some time by the ring before extraction is possible. The NP scenario only performs poorly in this case, since excitations in all likelihood decay before being extracted. Indeed, the relative power output of ratcheting over NP, shown in the left panel of Fig.~\ref{fig4}, therefore rises to as high as a factor of 35. The right panel of Fig.~\ref{fig4} demonstrates the importance of ratcheting over dark state protection alone, with the ratchet model delivering up to 20~\% better performance than FD.

In Fig.~\ref{fig4} we can see that advantage of ratcheting persists into a region where there is only a moderate bottleneck, i.e. up to $\gamma_x\sim~10^{-5}~\mathrm{eV} = 10 \gamma_o $. Referring to Fig.~\ref{fig3} we see that the ratchet power output in this region is already close to its maximum. 
Even so, if the extraction rate was arbitrarily tunable, then even a moderate bottleneck could be avoided completely and the main advantage of ratcheting removed. However, in photosynthesis, creating a fast extraction rate carries with it a clear resource cost: whereas antenna systems can be comparatively `cheap', reaction centres carry a much larger spatial footprint, being typically embedded in membranes and requiring significant surrounding infrastructure (such as concentration gradients produced by proton pumps).
The severity of the bottleneck is likely to be inversely proportional to the number of reaction centres.  A photocell design exploiting ratcheting in the moderate bottleneck regime would be likely then to generate optimal power per unit volume of material --- and similar design principles will apply to artificially designed molecular light harvesting systems. 

The choice of interaction strength $S$ is also important. Ratcheting achieves optimal results for $S \sim0.05$~eV in the bottlenecked region.
This dependence arises because the size of $S$ determines the gap between bright and dark or ratchet states. In turn, this controls the effective rate for `upwards' phonon-assisted transitions within excitation bands, as those rely on the absorption of a phonon and are proportional to $N(\omega_\Pi, T_p)$. Consequently, larger values of $S$ entail more directed dissipation into the lower states of each band, boosting the occupation of ratchet states. However at the same time, increasing $S$ leads to a lower ratchet state energy and so a lower trap energy -- and hence to a voltage drop. The trade off between these two competing influences leads to a maximum in ratchet performance. 

As we discuss in detail in the SI, ratcheting continues to convey an advantage in the presence of moderate levels of various real-world imperfections, such as site energy disorder, non-radiative recombination, and exciton-exciton annihilation~\cite{supinf}.

{\it Conclusion --} We have investigated the light harvesting properties of coupled ring structures, inspired by the molecular rings that serve as antennae in photosynthesis. Considering a vibrational as well as an electromagnetic environment allows the system to explore the full Hilbert space rather than just the restricted subset of Dicke ladder states. We have shown that the off-ladder states possess interesting and desirable properties, which can be harnessed for enhancing both the current and power of a ring-based photocell device. 

Several possible systems~\cite{supinf} could be used to observe the effect, from superconducting qubits~\cite{wallraff14} to macrocyclic molecules~\cite{anderson11}. Our approach generalises existing concepts for dark state protection~\cite{creatore2013, zhang2015} to arbitrary numbers of sites and importantly includes multi-exciton states, which introduce the ratcheting effect as a distinct additional mechanism for enhancing the overall light-harvesting performance. The optical ratchet enhancement is particularly well suited to situations where exciton extraction and conversion represent the bottleneck of a photocell cycle. 

In future work it would be interesting to explore combining optical ratcheting with enhancements of the primary absorption process, for example by exploiting the phenomena of stimulated absorption~\cite{kavokin2010} or superabsorption~\cite{higgins2014}.

\begin{acknowledgements}
The authors thank William B. Brown for insightful discussions. This work was supported by the EPSRC and the Leverhulme Trust. BWL thanks the Royal Society for a University Research Fellowship. EMG acknowledges support from the Royal Society of Edinburgh and the Scottish Government.
\end{acknowledgements}


\pagebreak

\begin{center}
\textbf{\large Supplemental Information: Quantum-enhanced capture of photons using optical ratchet states}
\end{center}

\setcounter{equation}{0}
\setcounter{figure}{0}
\setcounter{table}{0}
\setcounter{page}{1}
\makeatletter
\renewcommand{\theequation}{S\arabic{equation}}
\renewcommand{\thefigure}{S\arabic{figure}}
\renewcommand{\bibnumfmt}[1]{[S#1]}
\renewcommand{\citenumfont}[1]{S#1}

\section{Ring structure diagonalisation}

The Jordan-Wigner transformation maps a Pauli spin 1/2 system onto a `hard-core' boson model \cite{jordan1928}. This leads to a bosonic description of the collective excitons whilst maintaining Pauli's exclusion principle, which forbids double excitation of a single site. In our case each `spin' represents one of $N$ identical optical emitters / absorbers and the spin's `up' / `down' projection along the $z$-axis denotes the presence / absence of an exciton on the respective site. We assume a ring-like geometrical arrangement as shown in Fig.~1a of the main paper. Considering a ring rather than a spin chain with free ends introduces the additional complexity of an alternating periodic or anti-periodic boundary condition depending on the number of excitations~\cite{tokihiro1993}.

The Jordan-Wigner transformation is achieved by introducing bosonic creation and annihilation operators, which are defined in terms of the Pauli spin raising and lowering operators as follows:
\begin{eqnarray}
&{c}^{\dagger}_{l}=  {\sigma}^{+}_{l}e^{i \pi \sum_{j=1}^{l-1} {\sigma}^{+}_{j}{\sigma}^{-}_{j}} \, ,  \label{eq:JW1} \\
&{c}_{l}=  e^{-i \pi \sum^{l-1}_{j=1} {\sigma}^{+}_{j} {\sigma}^{-}_{j}} {\sigma}_{l}^{-} \, . \label{eq:JW2} 
\end{eqnarray}
Here, the $e^{i \pi \sum_{1}^{n-1} {\sigma}^{+}_{j} {\sigma}^{-}_{j}}$ factors are traditionally referred to as `strings', and their inclusion is necessary for producing the correct anti-commutation relation for fermions, which also apply to hard-core bosons:
\begin{eqnarray}\label{eq:commuation}
&\{{c}_{\alpha}, {c}^{\dagger}_{\beta}\} &= \delta_{\alpha, \beta} \, , \\
&\{{c}^{\dagger}_{\alpha}, {c}^{\dagger}_{\beta}\} &= \{{c}_{\alpha}, {c}_{\beta}\}= 0  \, .
\end{eqnarray}
Performing the transform (\ref{eq:JW1}, \ref{eq:JW2}) to the Hamiltonian of the main paper given by Eq.~(1) yields:
\begin{eqnarray}\label{eq:Ham1}
H_{s} & =  & \omega \hat{n} - S \sum_{j=1}^{N}({c}_{j}^{\dagger} {c}_{j+1} + {c}_{j+1}^{\dagger} {c}_{j} ) \nonumber \\
& + & S(e^{i \pi \hat{n}} +1)({c}_{N}^{\dagger} {c}_{1} + {c}_{1}^{\dagger} {c}_{N} ) \, ,
\end{eqnarray} 
where $\hat{n}=\sum_{j=1}^{N} {c}^{\dagger}_{j} {c}_{j}~=  \sum_{j=1}^{N} \sigma^{+}_{j} \sigma^{-}_{j}$ counts the number of excitons on the ring. The final term in Eq.~(\ref{eq:Ham1}) has been introduced to implement the required alternating boundary condition~\cite{tokihiro1993}. Note that this elegant and compact form of the Hamiltonian double-counts the interaction term for $N=2$, but this is of no concern here since we are only interested in $N \geq 4$.

Clearly, the operator $e^{i \pi \hat{n}} $ commutes with the Hamiltonian~(\ref{eq:Ham1}), so that the eigenvalues of $e^{i \pi \hat{n}} $ are good quantum numbers. The system can thus be partitioned into two parity subspaces of even and odd exciton number $n$, each of which can be solved separately. We complete the diagonalisation with the help of the Fourier transform

\begin{eqnarray}\label{eq:DFT}
&c^{\dagger}_{k}=  \frac{1}{\sqrt{N}}\sum_{j=1}^{N} e^{i k_j}c^{\dagger}_{j} \, , \\
&c_{k}=   \frac{1}{\sqrt{N}}\sum_{j=1}^{N} e^{-i k_j}c_{j} \, ,
\end{eqnarray}
where the $k_j$ are phase factors defined by   
\begin{eqnarray}\label{eq:Kseries}
&k_{j} = \frac{2 \pi}{N} j \, \, (n~\rm{odd}) \, , \\
&k_{j} = \frac{\pi(2j +1)}{N} \,\, (n~\rm{even}) \, .
\end{eqnarray}

\begin{figure*}[t]
$
\begin{array}{ccc}
  \subfigure[]{\includegraphics[width=2in]{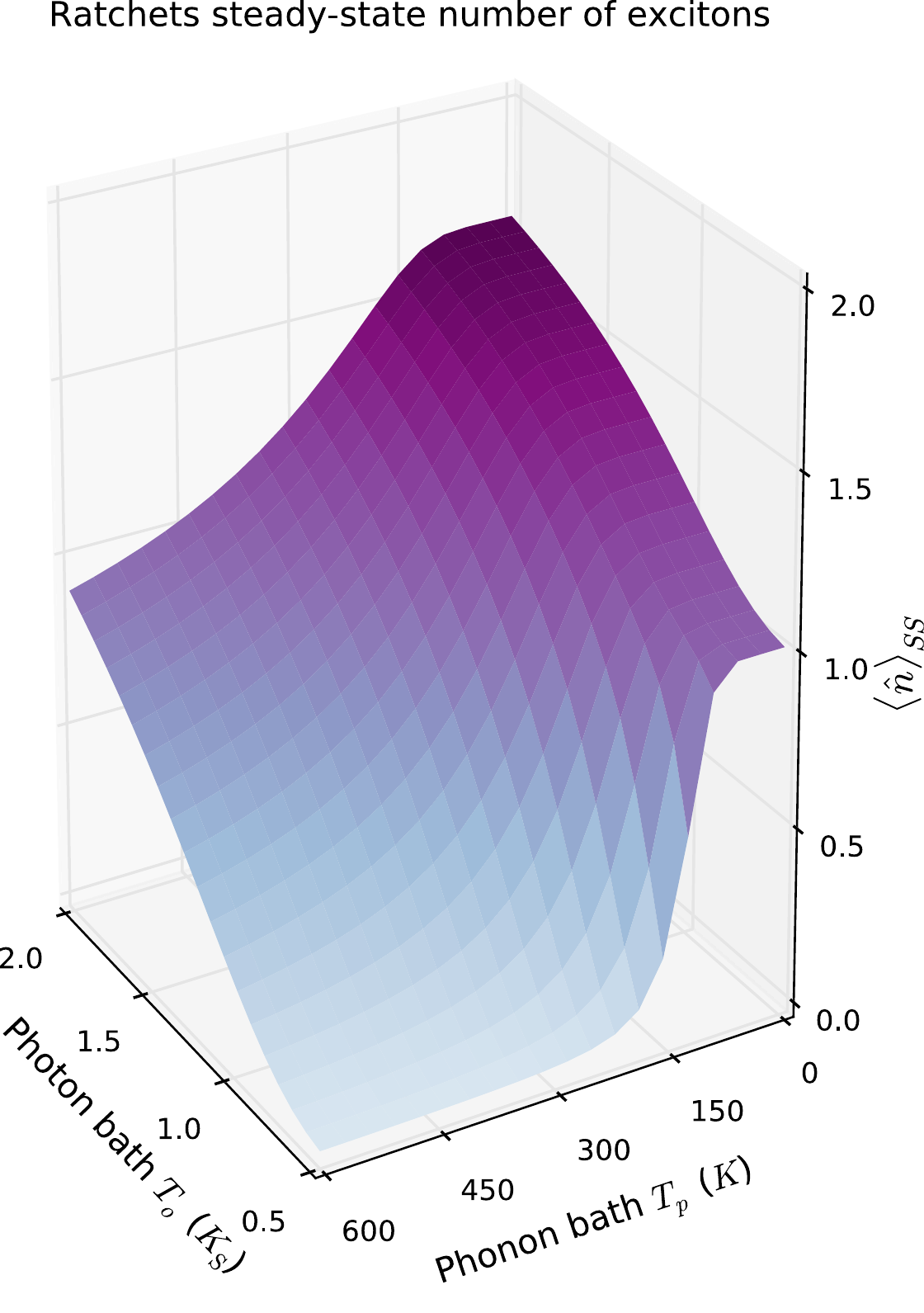}} &
  \subfigure[]{\includegraphics[width=2in]{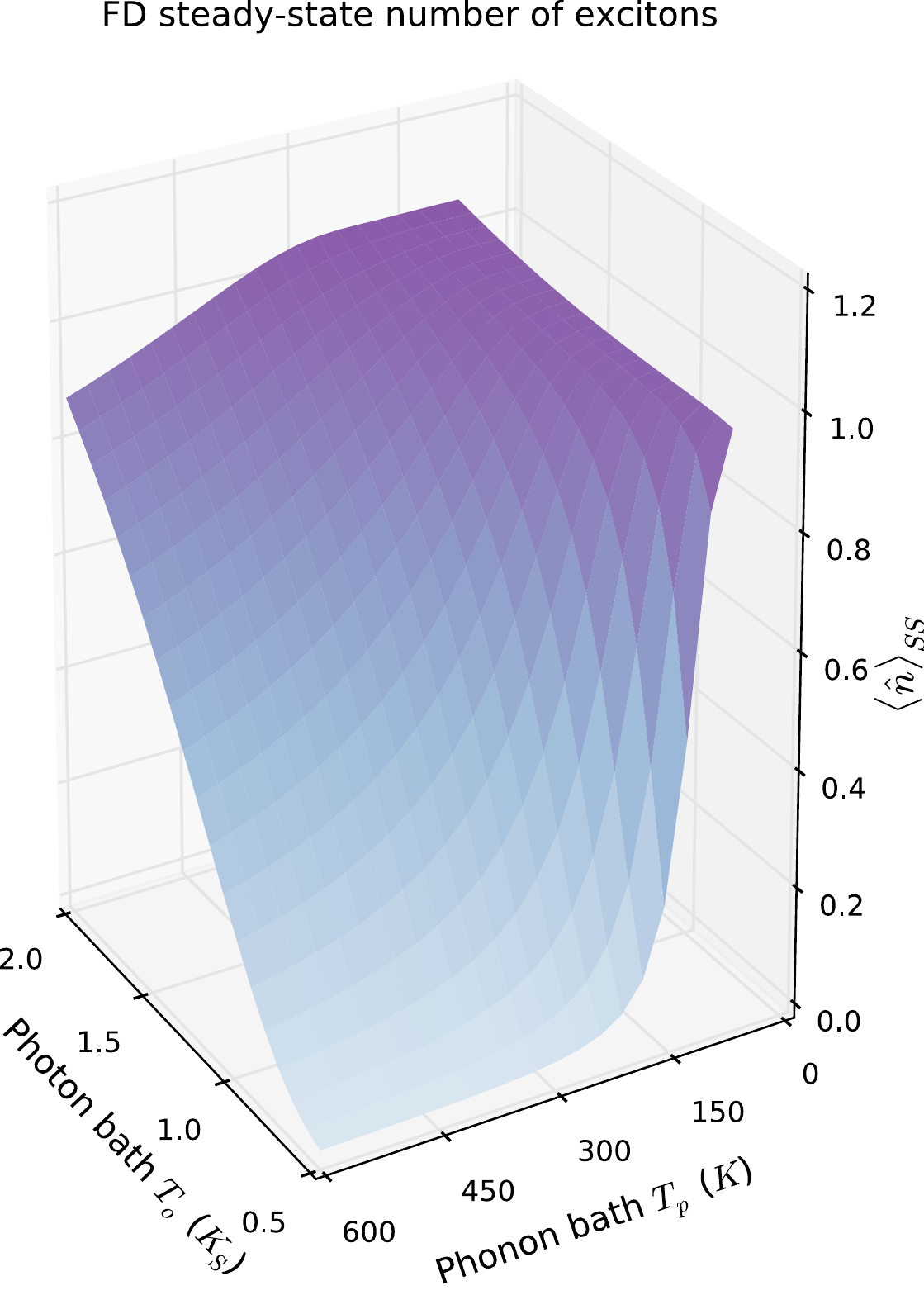}}&  
\subfigure[]{\includegraphics[width=2in]{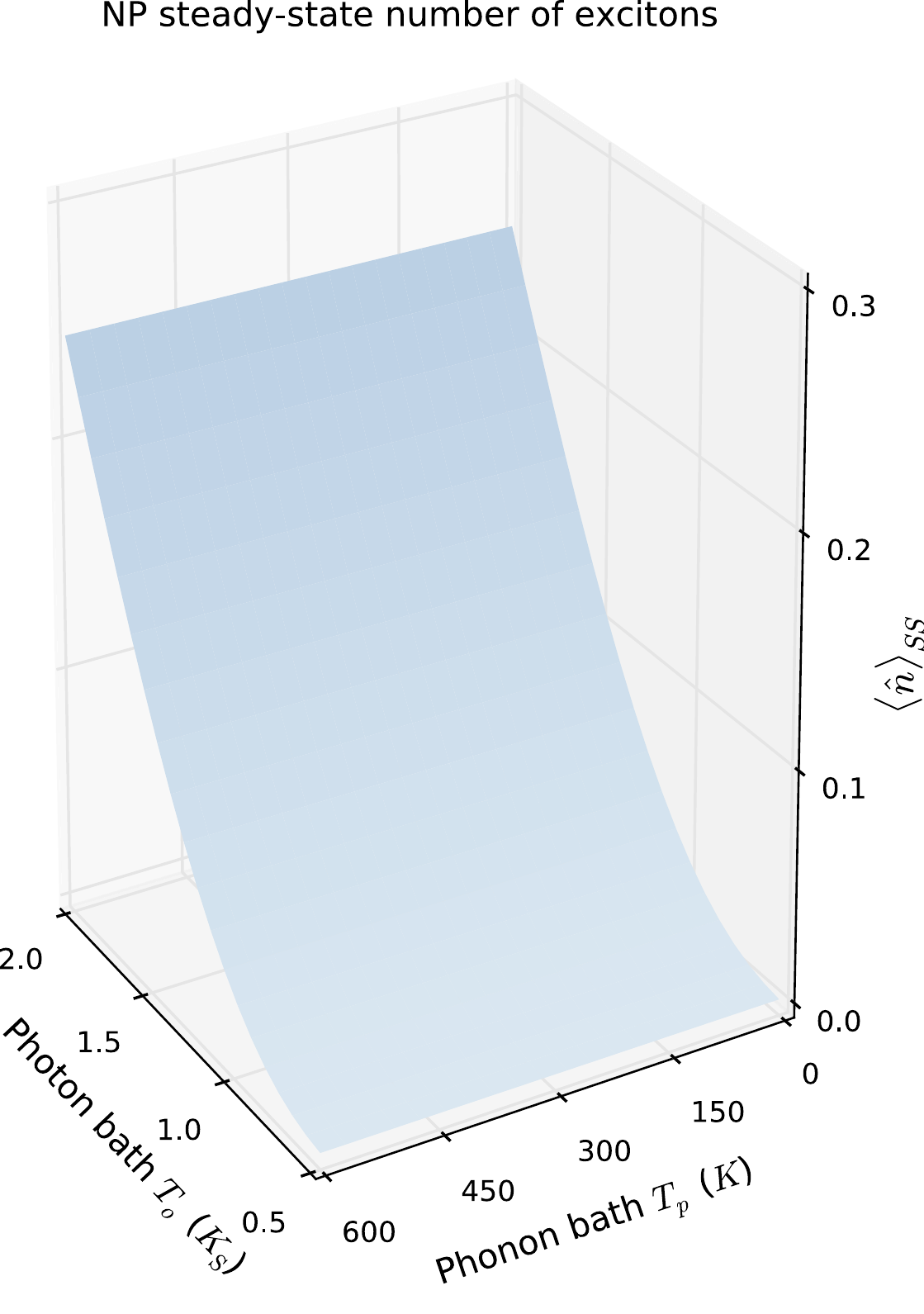}}  
\end{array}$
\caption{Number of excitons in the ring molecule in the steady state, with no attached trap. (a) Ratchets model; (b) forced dark (FD) model; (c) no phonon (NP) model. All three panels use the same colour mapping for ease of direct visual comparison. Panel (a) plots the same data also displayed in Fig.~2a in the main text. Parameters are from Table I in the main text.}
\label{fig:steadyex}
\end{figure*}

Each of the above $N$ distinct $k$-values corresponds to a single exciton state of the system. The eigenstates of the full multi-exciton system are given by combinations of these states.  The eigenstates are thus defined by sets $K$ whose elements are the $k$'s, and where the number of elements in $K$ represents the number of excitons in the state ($\left|{K}\right| = n$). The Pauli exclusion principle manifests itself by allowing only  different single exciton states to form a multiple exciton state. The eigenvalues are then given by \cite{tokihiro1993}
\begin{equation}\label{eq:lambdaK}
\lambda_{K} = \sum_{k \in K} (\omega -2S \cos{k} )
\end{equation}
with corresponding eigenstates
\begin{equation}\label{eq:gen}
\ket{K} = c^{\dagger}_{k_{1}}... c^{\dagger}_{k_{n}}\ket{0} \, , 
\end{equation}
where $\ket{0}$ is the ground state. Obviously, Eq.~(\ref{eq:gen}) yields an eigenstate with $n$ excitations and each band comprises a total of $\frac{N!}{n!(N-n)!}$ possible states. In its diagonalised form the Hamiltonian Eq.~(\ref{eq:Ham1}) is then simply given by
\begin{equation}\label{eq:Ham}
H_{s} =\sum_{K} \lambda_{K} \, \ket{ K} \bra{K} \, .
\end{equation}

\section{Optical transiton matrix elements}

The optical transition operator $\hat{J}^{\pm} = \sum_{i=1}^{N} \hat{\sigma}^{\pm}_{i}$ connects eigenstates differing by one exciton. The transition rates are proportional to $\Gamma_{K,K'} = \abs{ \bra{K'}J_{+} \ket{K}}^{2} $ and can be explicitly evaluated with the help of Slater determinants, albeit involving some tedious calculations \cite{tokihiro1993,spano1991}, yielding:

\begin{align}\label{eqn:matrix_element}
\Gamma_{K,K'} = \bigg\vert& 2^{n}N^{\left(-n +\frac{1}{2}\right)} \delta \left(\sum_{j}k'_{j} , \sum_{i}k_{i} \right)  \\ \nonumber
&\frac{\Pi_{i>i'}(e^{i k_{i}} - e^{i k_{i'}}) \Pi_{j>j'}(e^{-i k'_{j}} - e^{-i k'_{j'}})}{\Pi_{i=1}^{n} \Pi_{j=1}^{n+1}(1-e^{i(k'_{j}-k_{j} ) })}\bigg\vert^{2} \,,
\end{align}
where $\delta(x,y )$ is 1 if $x= y + 2 \pi m$ for integer $m$ and zero otherwise. The indices $i,i',j,j'$ inside the sums and products of the above expression refer to the elements of unique pairs of $k$-values composing $K$ or $K'$~(see~\cite{tokihiro1993} for full details).

\section{Steady state exciton populations}

In Fig.~2 of the main text we display the steady-state exciton population in the ratchets model as a function of phonon and photon temperatures, and also the relative population of the ratchets model over the forced dark (FD) scenario. To complete the picture, in Fig.~\ref{fig:steadyex} we show the absolute values of the steady-state exciton count for all three scenarios. In the no phonons (NP) case, as expected there is no dependence on the phonon temperature. Indeed, since there is no dark-state protection in the NP case, we find a lower exciton population than in the other two cases; the difference is more pronounced at lower phonon bath temperature, when dark states are most likely to be populated. Naively, one might expect the surface for the FD scenario to plateau at one, however, the ladder of bright states remains available in this case and via it population can access and get trapped in higher lying dark states above the single exciton band. We expect this to happen for high optical and low but finite phonon temperature, and indeed in this region the surface rises above one. Note that unlike panels (a) and (c) in  Fig.~\ref{fig:steadyex}, (b) starts at a phonon bath temperature of $T=75$~K due to numerical instabilities in our steady-state solver. At much lower phonon temperature, where phonon-assisted relaxation becomes effectively uni-directional, we expect the FD surface to drop towards one, as it gets increasingly unlikely for population to bypass or escape the single excitation subspace. Using alternative methods, we have verified this is indeed the case (not shown).

\begin{figure}[!]
\centering
\hspace{0cm}\includegraphics[width=\linewidth]{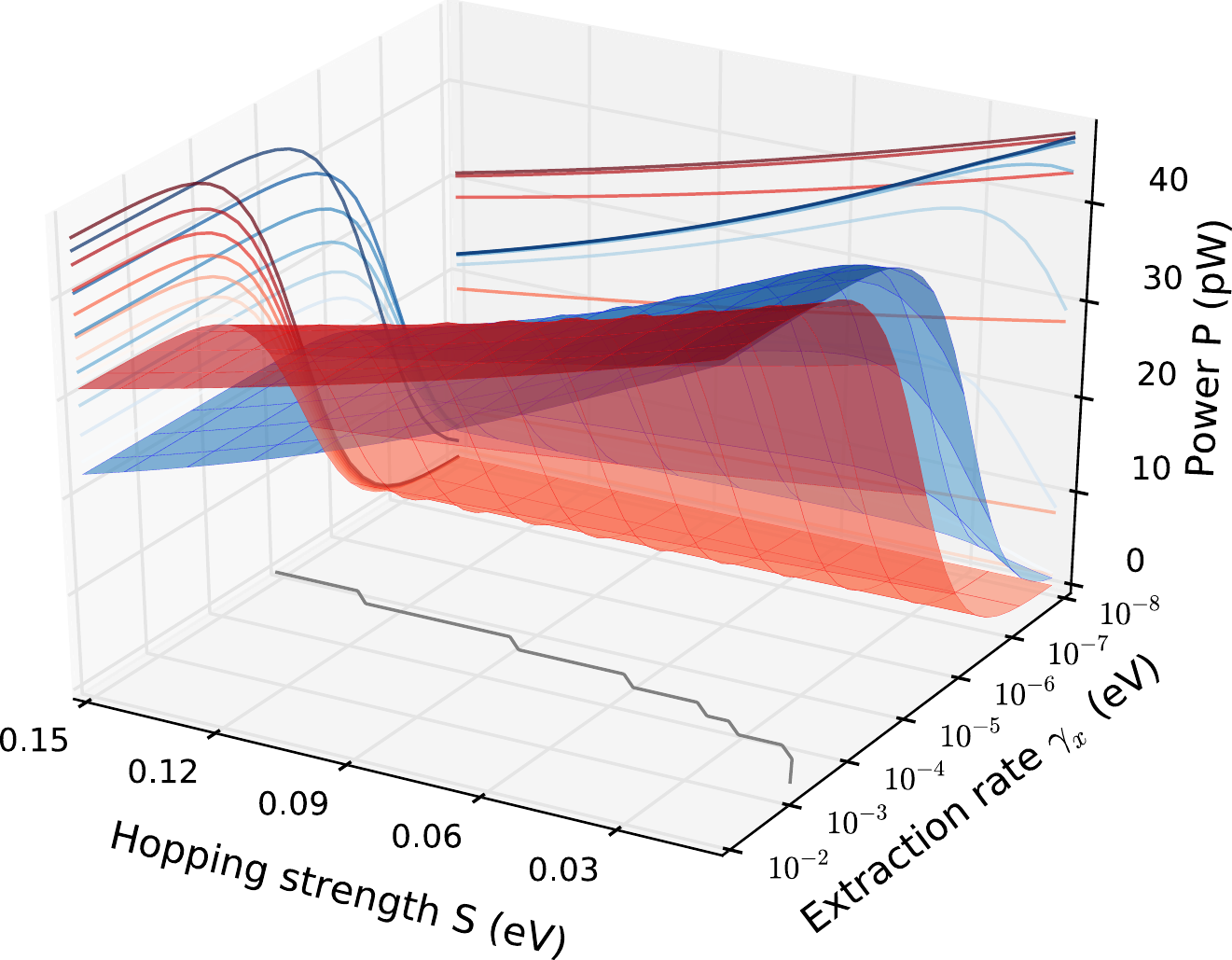}
\caption{Absolute power output in the ratcheting (blue surface) and FD (red surface) cases as a function of (single site) extraction rate and F\"orster coupling. Here the collective extraction model is used. Other parameters are from Table I and thus identical to those of Fig.~4 of the main text. The sets of curves projected onto the side walls of the 3D plot are cuts through the data for fixed values of the appropriate parameter. For each point, the optimal trap decay rate is found by numerical search. The two surfaces cross in the region where the bottleneck is lifted, with ratcheting generating more power for a severe and intermediate bottleneck. The line at which the curves cross is projected onto the bottom ($xy$) plane of the figure.}
 \label{fig:collabs}
\end{figure}

\begin{figure}[!]
\centering
\includegraphics[width=\linewidth]{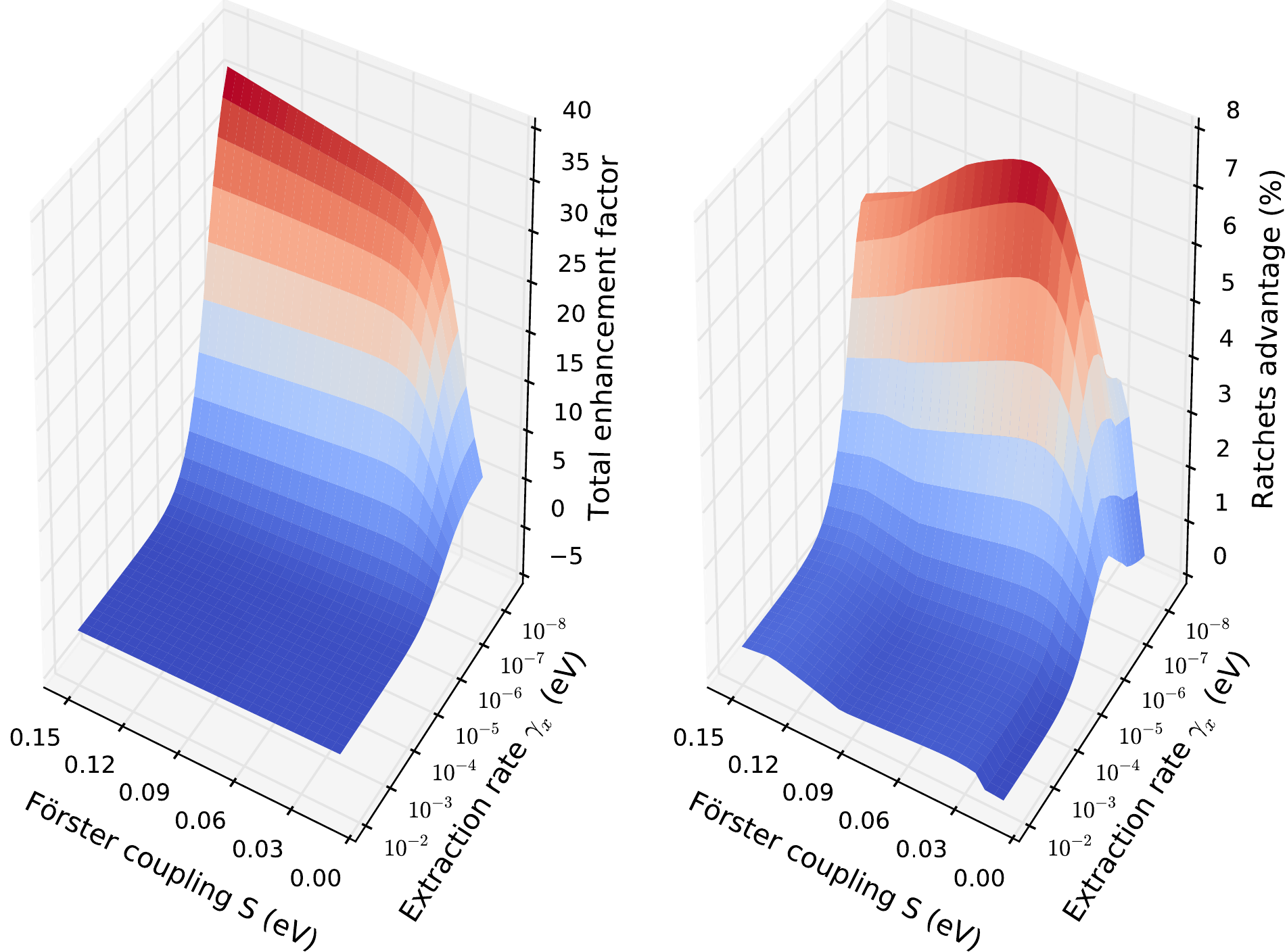}
\caption{ \label{fig:bottleneck} For the collective extraction model, enhancement of power output for ratchet states over NP (left, displayed as a multiplicative factor) and FD (right, displayed as a percentage), as a function of both hopping strength $S$ and extraction rate $\gamma_x$. The simulations use the same model and parameters as Fig.~\ref{fig:collabs}.
}
 \label{fig:collrel}
\end{figure}

\section{Extraction mechanisms}
\label{sec:ext}

Our photocell cycle relies on the extraction of the excitons from the ring to the trap site. Below we give an explicit description of two different mechanisms and models, one of which features in the main text, the other whose results are displayed for comparison in this supplement.

For the incoherent hopping process that we use in the main text we assume that excitons are extracted incoherently from a single site of the ring to the trap. We then simply include a dissipator:  
\begin{equation}\label{inc_ trap}
D_{x}[\rho]  = \gamma_{x} {\cal D}[\sigma^{-}_1 \sigma^{+}_{t}, \rho] \,.
\end{equation}
For optimal extraction the trap's energy is resonant with the desired extraction level, i.e.~for the ratcheting and FD cases, the lowest state in the first exciton band, $\omega_{t} = \omega - 2S$; for the NP case, the trap is instead made resonant with the bright state, $\omega_{t} = \omega + 2S$. For consistency and a fair comparison across the different mechanisms, the energy of the trap site is also adjusted in the same way for the extraction mechanism described next.

We use the single site extraction process in the paper since it is relatively straightforward to implement. However, it is not necessarily the most efficient --- so in this supplement we will also consider the possibility of extraction directly from particular eigenstates of the ring. This can sometimes be achieved by suitable positioning of the trap: e.g.~placing the trap at the ring's centre with equal coupling to all other sites allows extraction from symmetric states~\cite{higgins2014}. Unfortunately, this does not straightforwardly generalise to the more complex anti-symmetric dark states. For the special cases of $N=2,3$ alternating phases on the interaction terms achieves the desired coupling \cite{creatore2013, zhang2015}; however, scaling up to more sites would require increasingly exotic molecular orbitals. We therefore implement a collective state extraction method using a set of operationally motivated Lindblad operators $\hat{C}_i$. In the case of the ratchets and the FD model, the $\hat{C}_i$ move population from the lowest state of a particular band of $\hat{H}_s$ into the lowest state in the exciton band below, exciting the trap in the process. In the NP case, the $\hat{C}_i$ go from the highest state of one band to the highest of the band below (since these are the only states accessed in this model). We incorporate this process into our model through the dissipator
\begin{equation}\label{Lind_trap}
D_{x, \textrm{col}}[\rho] =  \gamma_{x} \sum_{i} {\cal D}[\hat{C}_{i}, \rho] \,,
\end{equation}
where, once again, $\gamma_{x}$ is the effective extraction rate. Once on the trap, excitons are assumed to undergo charge separation at rate $\gamma_{t}$, which we model as an irreversible decay process. 
\begin{figure}
\includegraphics[width=\linewidth]{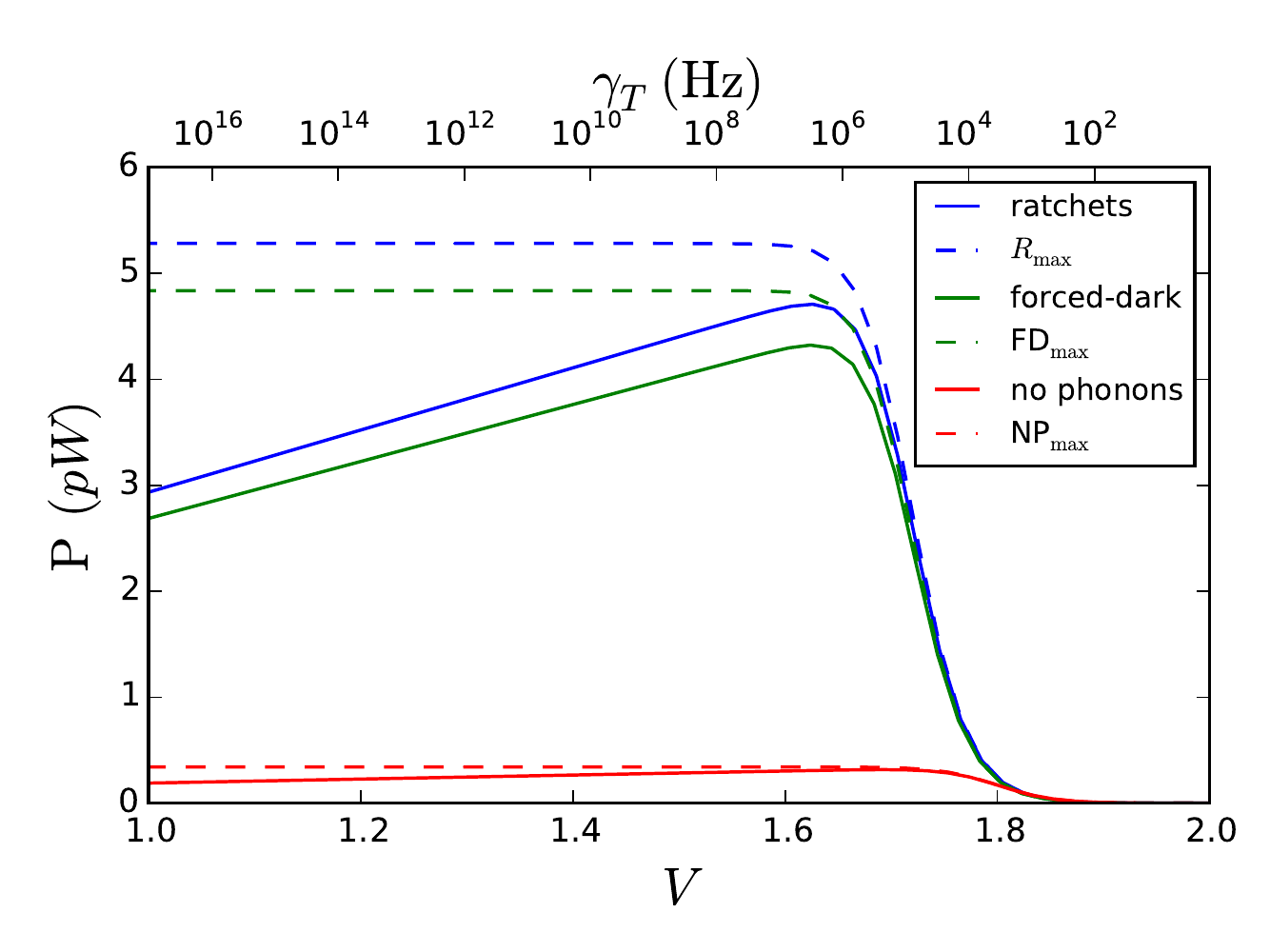}
\caption{ \label{fig:fig3old} Power (solid) and its limit (dashed) as a function of voltage based on single site incoherent Lindblad extraction. Three scenarios are shown, ratcheting (blue), forced-dark (FD, green) and NP (red). The power limit is given by the maximum possible voltage multiplied by the current $P_{max}=I(\omega+2S)$ \cite{creatore2013}. Model parameters are from Table I of the main text. Each plot is found by varying $\gamma_{t}$ independently for the different models. The upper $x$-axis shows how $\gamma_{t}$ varies in the ratchets model over many orders of magnitude for relatively small changes on the voltage scale.}
\end{figure}

\begin{figure*}[!]
\centering
\includegraphics[width=0.67\linewidth]{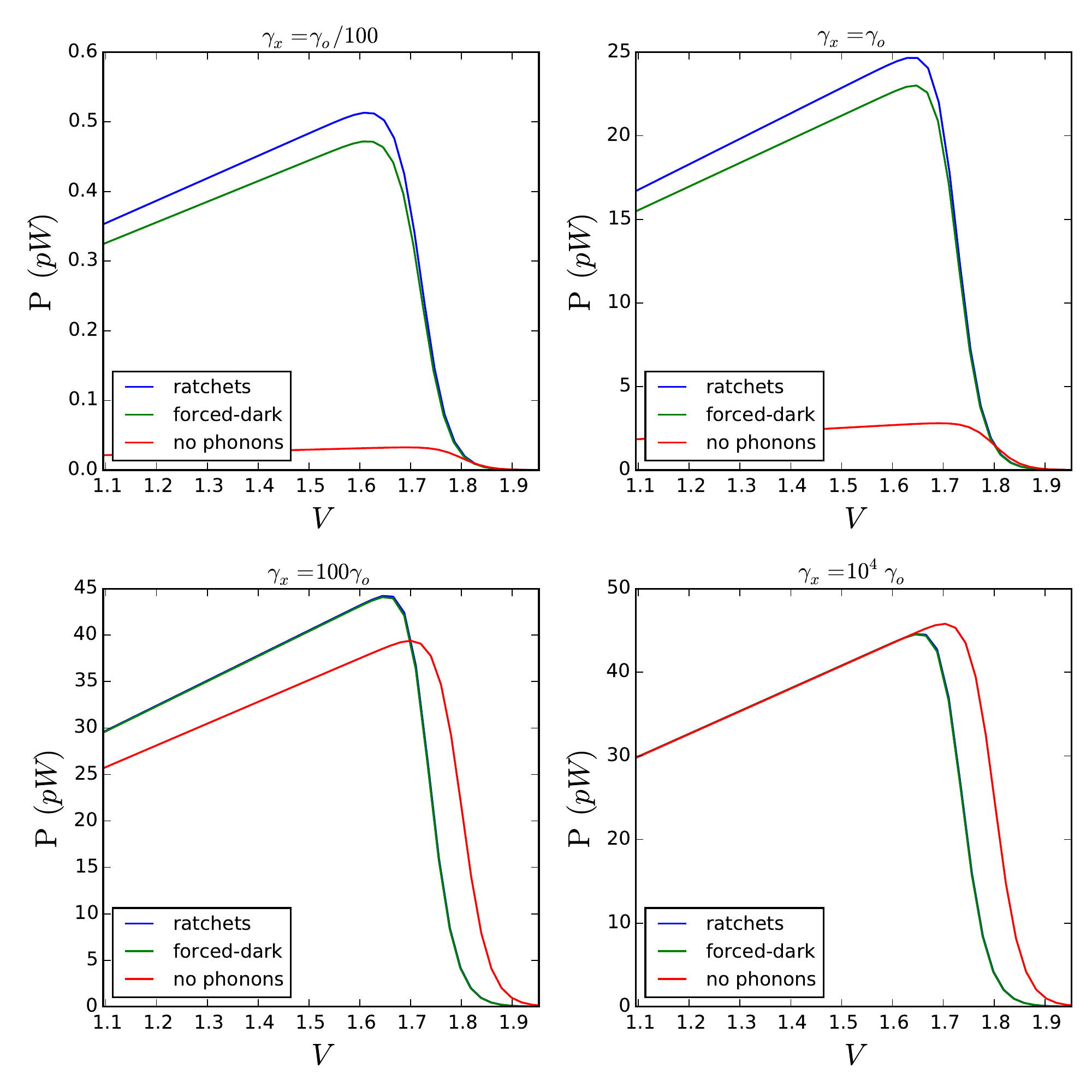}
\caption{ \label{fig:bottleneck} Power-Voltage characteristics for the single site incoherent extraction model. Parameters are as in Table I of the main paper except for $\gamma_x$. Each panel compares the three models (ratchets, NP, FD) for a different extraction rate $\gamma_x$. The advantage of ratcheting is most apparent for slower extraction rates, i.e.~in the bottleneck region.}
\end{figure*}

Fig.~\ref{fig:collabs} shows the absolute power for the ratchets and NP model for the collective trapping model. This plot is equivalent to Fig.~3 of the main paper which displayed the same quantities for the single site incoherent trapping model. Similarly, Fig.~\ref{fig:collrel} shows the {\it relative} power enhancement of ratcheting over both artificial models for the collective trapping model --- again this is the equivalent plot to Fig.~4 in the main paper but here for the collective extraction model. Unsurprisingly, the absolute power extracted in the collective model is larger than the incoherent model. This is because we are here extracting directly from the ratchet or dark states and thus have access to the `entire' exciton rather than just its projection onto a single site of the ring. The relative advantage of ratcheting over FD has a similar form for both extraction mechanisms, though it is smaller for the collective model.

\section{Power-Voltage curves}

In Figs.~3 and 4 of the main paper we are displaying maximum output power as a function of $S$ and $\gamma_x$. Each point in those plots is a result of generating a power-voltage curve for each choice of parameter, and finding the maximum power. Each such curve is generated by varying the trap decay rate $\gamma_t$.  We show examples of these power-voltage curves in Fig.~\ref{fig:fig3old}, considering the usual three scenarios: ratchets, FD and NP.  On the upper $x$-axis we also show the values of $\gamma_t$ corresponding to each voltage taking the example of the  ratchets case; as shown $\gamma_t$ varies over orders of magnitude for only modest voltage shifts. These curves make use of the parameters specified in Table I of the main text, and a comparison with Figs.~3 and 4 of the main paper shows that $S=0.02$~eV and $\gamma_x = 10^{-7}$ eV are in fact not optimal for showing the full ratcheting advantage.

Further examples of $P-V$ curves are shown in Fig. \ref{fig:bottleneck}, each for different single site incoherent extraction rates, fixing $S=0.02$~eV. We can see here the advantage of ratcheting in the intermediate and severe bottleneck regions. For very fast extraction, any absorbed photon gets immediately transferred to the trap, so that neither dark-state protection nor ratcheting offer any advantage (the system is already operating as efficiently as possible). In this case NP achieves the same current as ratchets and FD, but ends up producing a higher resultant power as its trap transition frequency -- and thus effective voltage -- is higher, as explained in Sec.~\ref{sec:ext}. 

\section{Extraction Bottleneck 2D plots}

\begin{figure}[!]
\centering
\includegraphics[width=1\linewidth]{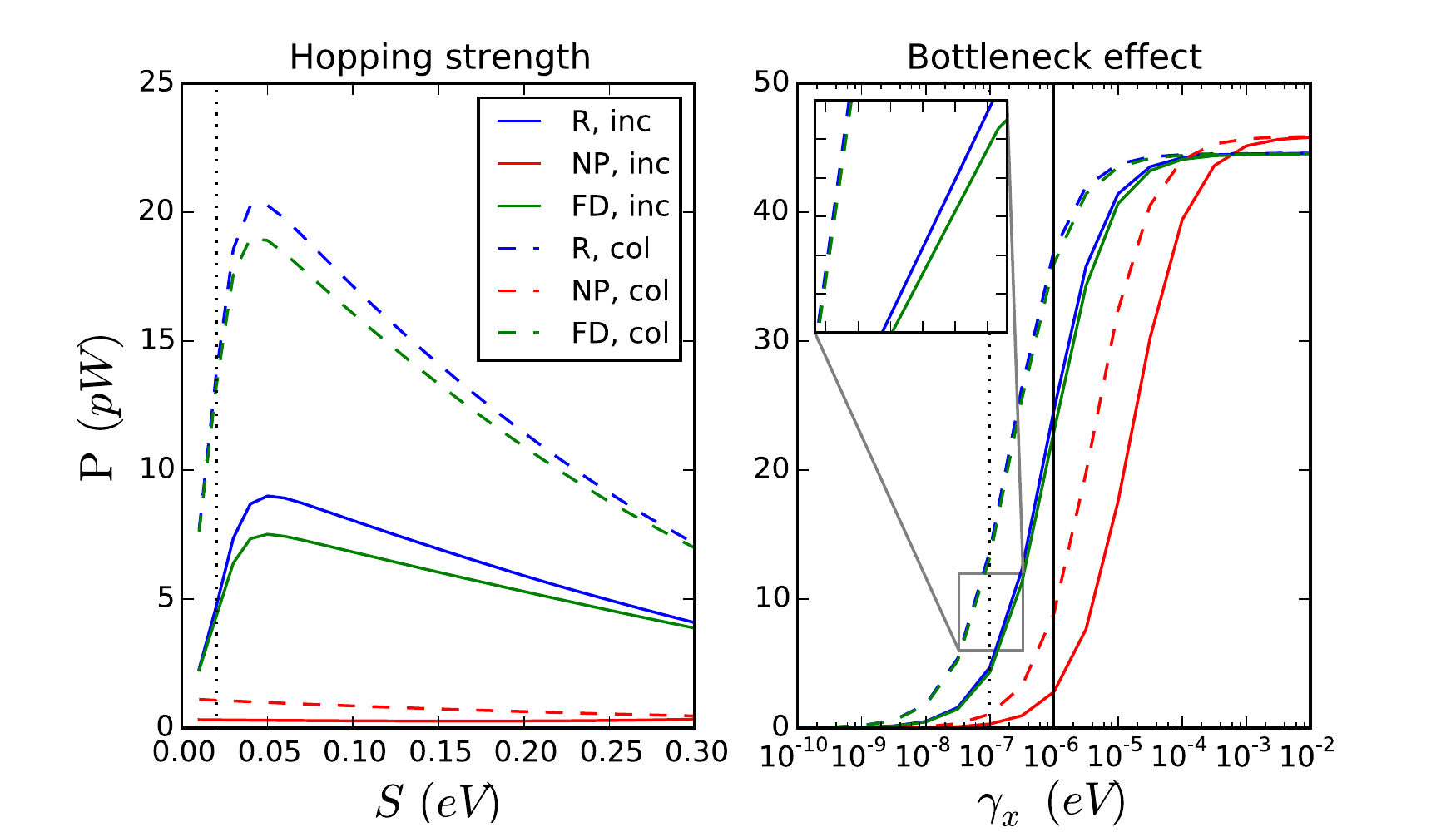}
\caption{Left: absolute output power as a function of $S$ (for fixed $\gamma_x=10^{-7}~$eV). Right: absolute output power as a function of $\gamma_x$ (for fixed $S=0.02~$eV). Curves for all three scenarios -- ratchets (R), NP, FD -- are displayed. The two extraction mechanisms of incoherent single site (inc) and incoherent collective (col) are also shown. All other parameters are as specified in Table I of the main text.}
\label{fig:2Dbottle}
\end{figure}

In Fig.~\ref{fig:2Dbottle} we show several 2D sections of the surfaces in Fig.~3 of the main paper and Fig.~\ref{fig:collabs} -- i.e.~we show absolute output power as a function of $S$ (for fixed $\gamma_x=10^{-7}~$eV) in the left panel and as a function of $\gamma_x$ (for fixed $S=0.02~$eV) in the right panel. These are displayed for all three scenarios (ratchets, FD, NP) and for both collective and incoherent extraction mechanisms.

\section{Imperfections}

\subsection{Disorder}
\label{sec:disorder}

To model disorder, we generate random site energies that are normally distributed around the transition frequency $\omega+ 2S = 1.8$~eV with standard deviation $\sigma$. The $I, V$ value pairs for each trap decay frequency $\gamma_t$ are averaged over 100 different implementations, and are based on the Bloch-Redfield dissipator~\cite{breuer02} for the optical and phonon interactions.

\begin{figure*}[!]
\centering
\includegraphics[width=0.4\linewidth]{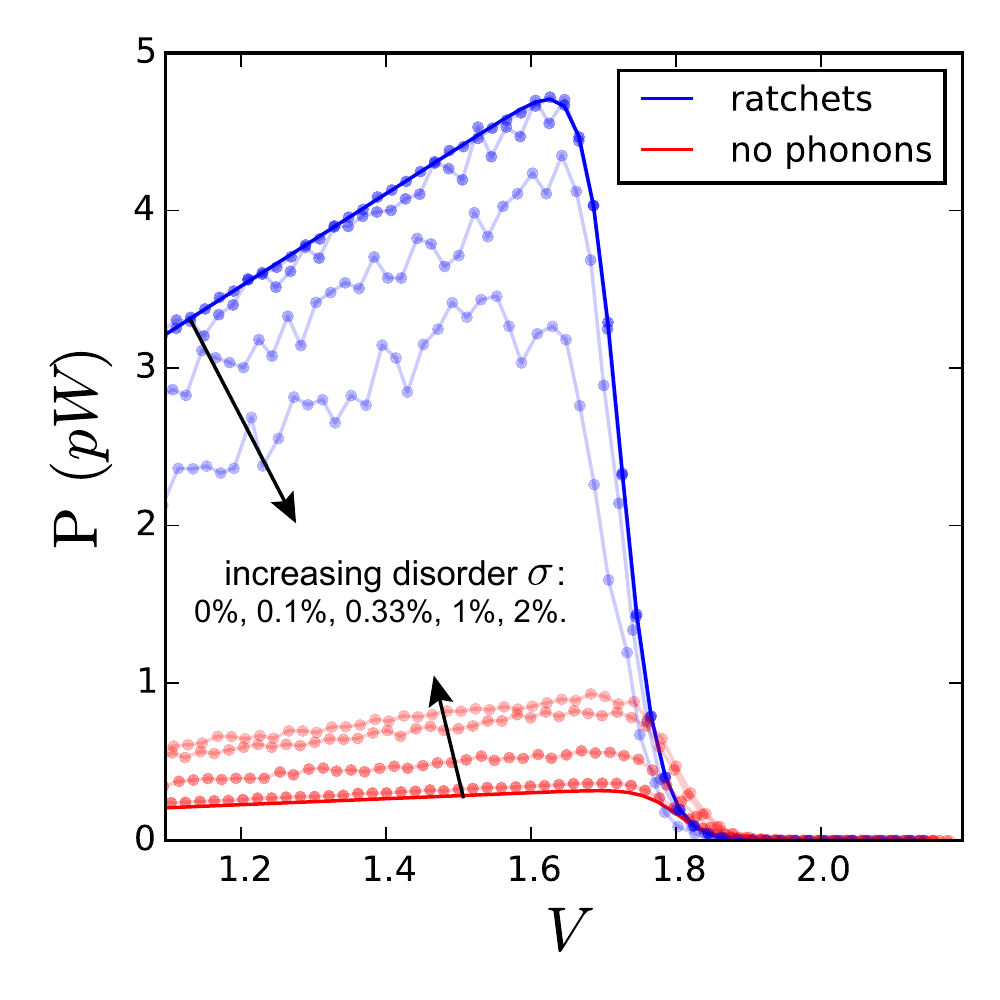}
\includegraphics[width=0.4\linewidth]{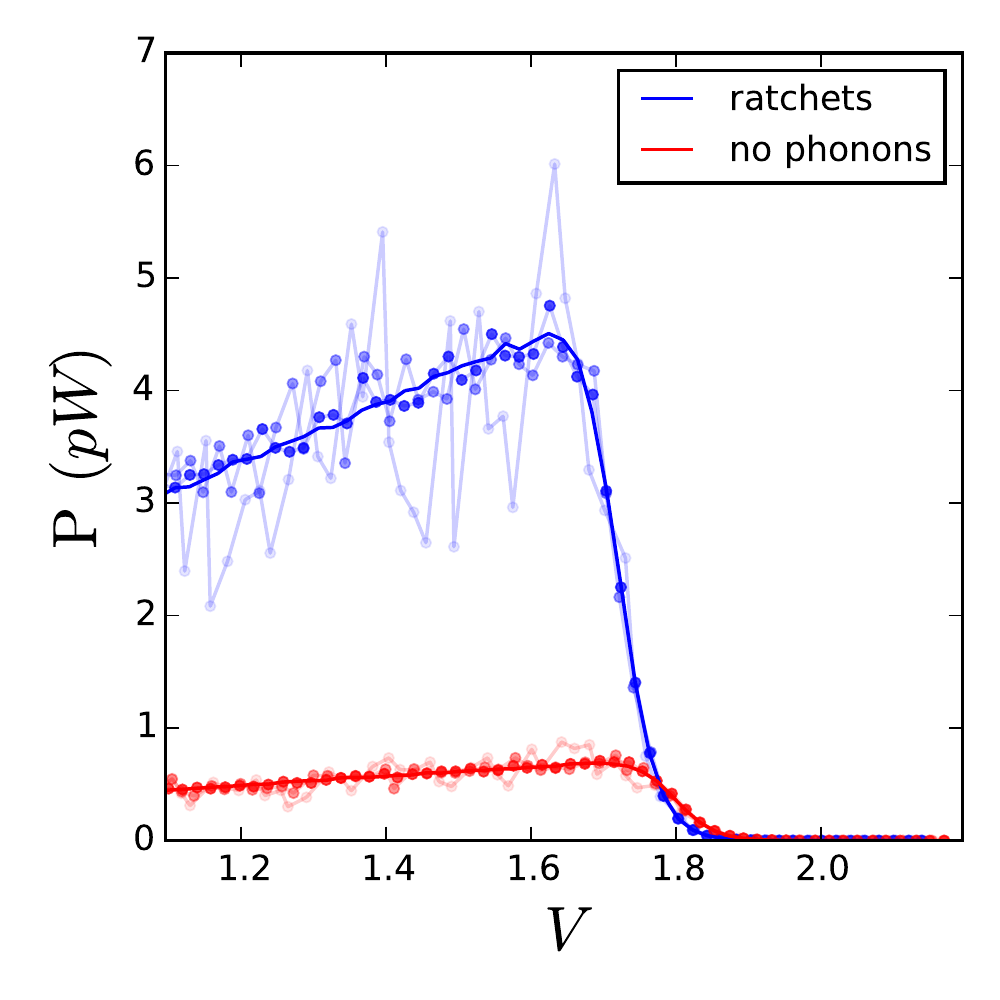}
\caption{ \label{fig:disorder} Effect of site energy disorder on the achievable power. {\bf Left:} The darkest blue and red curves represent the ideal case of no disorder, the lighter curves with dots correspond to random site energy fluctuations with a standard deviation from the idealised case of 0.1\%, 0.33\%, 1\%, and 2\%. All these curves are averaged over 100 different randomly selected configurations.  {\bf Right:} Convergence of disorder curves for $\sigma = S/2$ (corresponding to about 0.57\% disorder of site energies). The solid lines represent the average over 1000 runs, whereas the joined circles from faint to more opaque correspond to the average of 1, 10, and 100 runs. Note that certain random configurations prove to be highly efficient -- see text for a discussion. Model parameters are from Table I of the main text.}
\end{figure*}

Results for different degrees of disorder are shown in the left of Fig.~\ref{fig:disorder}. For small levels of disorder $\sigma < S$, there is little deviation from the ideal case, and 100 implementations are indeed enough to give fairly smooth curves. Once $\sigma$ approaches $S$ (0.02~eV corresponds to a $\sigma$ of around 1.14\%), however, not only is the achievable power with ratchets visibly reduced but our statistical ensemble is also not large enough to smooth out those larger performance variations, resulting in jagged-looking curves. 

Interestingly, the performance of the NP case improves with increasing disorder. This may be due to the fact that even in asymmetric circumstances eigenstates with a varied range of interesting optical properties may arise~\cite{fruchtman16} in the fully accessible Hilbert space, and the absorption dynamics is no longer restricted to only the single lowest ladder transition.

We do not show the FD scenario here, as the presence of disorder makes it impossible to properly distinguish between allowed `ladder'  and FD `forbidden' optical transitions. However, we note that random sampling for small levels of disorder (well below relative site energy fluctuations on the order of a percent) -- where the symmetries and properties of system eigenstates remain largely unperturbed -- indicates that FD always performs worse than the full ratchets model, as expected.

The right panel of Fig.~\ref{fig:disorder} shows the averaging from a single up to 1000 implementations of randomness for $\sigma = S/2$. Interestingly, the single run has some datapoints which can even exceed the ideal, fully ordered, ratcheting performance --- though the ensemble average power is clearly below that in the fully ordered case. We suspect these outliers correspond to highly efficient asymmetric configurations:  For the case of a dimer and dark-state protection it has recently been established that certain asymmetric arrangements can potentially outperform symmetric ones~\cite{fruchtman16}. However, the unravelling of the precise interplay of mechanisms at play will be more complicated in this larger system, but would be an interesting line of investigation for a future study.

Note that we keep our extraction and trap model unaltered for our disorder calculations, i.e.~we extract with a trap energy of $\omega_t = 1.8~\mathrm{eV} \pm 2S$ for the NP and ratchets case, respectively. This simplifying choice may in general entail slightly over- or underestimating the effective voltage of our trap. However, the relative voltage discrepancy arising from this assumption is expected to be limited by energy shifts due to disorder given by $\sigma$. As seen in both panels of Fig.~\ref{fig:disorder}, disorder in the model affects the power output for both ratchets and NP much more significantly than that, meaning our results are clearly dominated by other processes as opposed to just being an artefact of our simplified trapping model.

\subsection{Non-radiative decay}

\begin{figure*}[!]
\centering
\includegraphics[width=\linewidth]{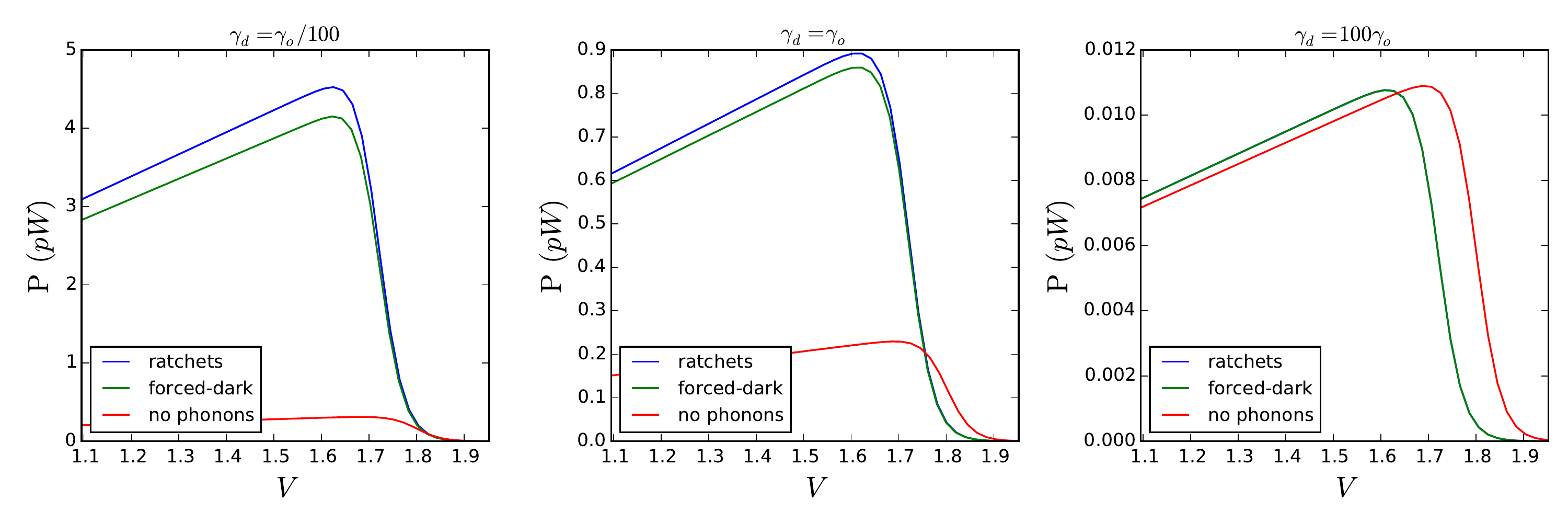}
\caption{ \label{fig:non-rad} Non-radiative decay processes acting independently on each site, with effective rates as shown above each panel. All other parameters are from Table I of the main text. Aggressive non-radiative recombination severely reduces the achievable power output of all three models, ratchets, NP, and FD, lending an edge to NP where excitations are never stored for long in excited levels. }
\end{figure*}

We model non-radiative decay by including a full set of uni-directional Lindblad lowering processes acting in the site basis (as opposed to the collective optical decay events which are present due to the interaction with the shared electromagnetic environment). 

Increasing non-radiative decay rates significantly decreases the achievable current and power output of the photocell as shown in Fig.~\ref{fig:non-rad}. Interestingly, non-radiative decay does not affect all three scenarios equally, and the ratcheting advantage diminishes with larger non-radiative recombination rate. The ratchets states enjoy no special protection against non-radiative decay acting on individual sites, and when they carry more than one excitation the impact of non-radiative decay is felt more acutely. However, the combined ratcheting and dark state protection advantage persists as long as non-radiative decay does not become the dominant loss channel. 

\subsection{Exciton-exciton annihilation}

\begin{figure*}[!]
\centering
\includegraphics[width=\linewidth]{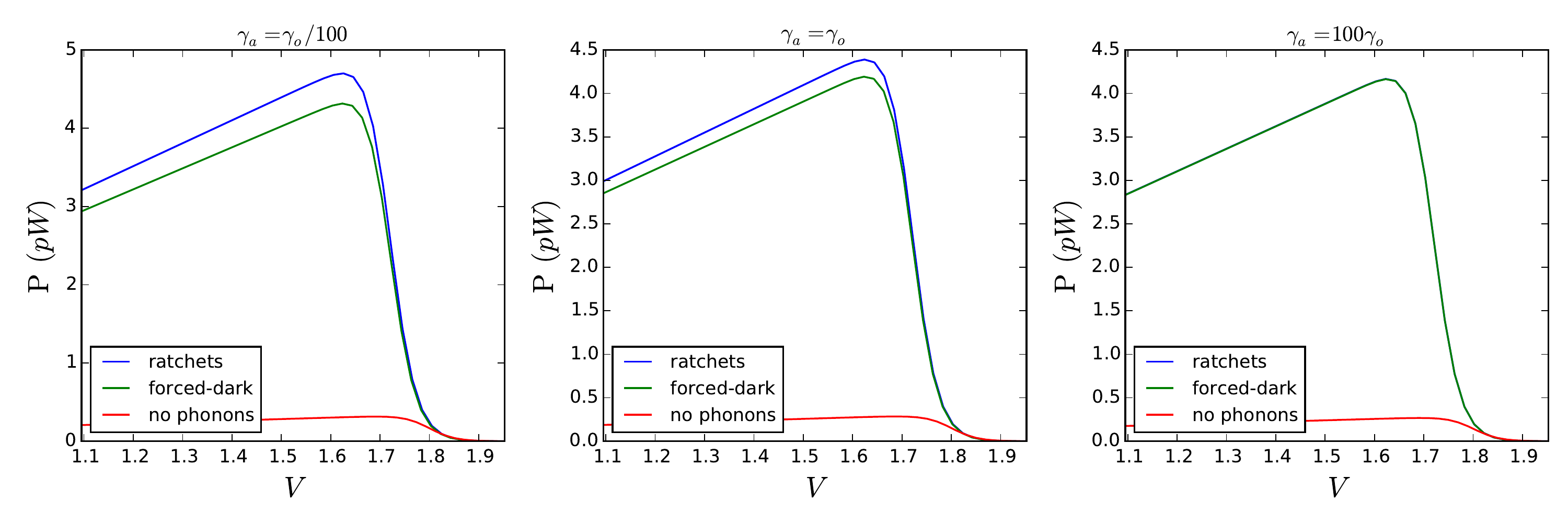}
\caption{ \label{fig:annihilation} The effect of exciton-exciton annihilation at an effective rate as shown above each panel. All other parameters are from Table I of the main text. As expected neither NP nor FD are much affected, and notably the ratcheting advantage only fully vanishes once the exciton-exciton annihilation rate exceeds the spontaneous emission rate by an order of magnitude or more. }
\end{figure*}

We now discuss the impact of exciton-exciton annihilation (EEA) on the ratcheting efficiency enhancement. EEA is the result of a two step dynamical process~\cite{may14}. First, excitons in adjacent sites on the ring can fuse together, creating an excitation on a single chromophore that has twice the energy of the first excited exciton state so far considered. The mechanism for this is, like exciton transfer, F\"orster coupling -- i.e. a dynamic dipole-dipole interaction between the adjacent excitations. 
The second step is fast non-radiative decay of the double energy state on the excited chromophore back down to the regular single exciton state. 

The Hamiltonian underlying the first step might be expressed by using an extension of the notation we have employed in the main paper. We now allow three levels per chromophore: the already-considered ground state $\ket{G}$ and low energy exciton state $\ket{E}$, as well as the newly introduced double energy excitation $\ket{D}$. With these state labels the Pauli operator we use in the main text is $\sigma^+ = \ket{E}\bra{G}$, and now we define a new spin raising operator $\eta^+ = \ket{D}\bra{E}$. The Hamiltonian then becomes:
\begin{align}\label{eq:Ham2}
H_{s'} = &\omega \sum_{i=1}^{N} (\sigma_{i}^{+} \sigma_{i}^{-}+2 \eta_{i}^{+} \eta_{i}^{-})+ \\
&S \sum_{i=1}^{N}(\sigma_{i}^{+} \sigma_{i+1}^{-} + \sigma_{i}^{+} \eta_{i+1}^{-} + H. c. \, ) ~, 
\end{align}
where we have made the simplifying assumption that the two excitation energy transfer has the same strength ($S$) as the single excitation one. This is reasonable, implying a similar transition dipole from $\ket{G}$ to $\ket{E}$ as from $\ket{E}$ to $\ket{D}$. By diagonalizing Eq.~(\ref{eq:Ham2}) for an $N=4$ system, in a basis of states $\{\ket{G}, \ket{E}, \ket{D}\}$ on each site, we are able to straightforwardly obtain the new eigenstates of the system. Again, since the Hamiltonian preserves total excitation number (assuming that $\ket{D}$ states count as two excitations), then the eigenstates fall into a series of bands,  each corresponding to a different excitation number. The zero excitation ground state and the four single excitation eigenstates are obviously unaffected by the $\ket{D}$ states, but the second band now consists of ten, not six, possible states, which are now combinations of both $\ket{D}$ states on single sites and $\ket{E}$ states on two sites. The lowest energy state in this second band (let us label it $\ket{\cal G}_2$) would be reached following rapid relaxation following ratchet action from the first band. We have calculated that $\ket{\cal G}_2$ contains only a 16\% $\ket{D}$ state character (by probability). Considering that population of the $\ket{D}$ state is crucial for EEA to occur, we can conclude that EEA will be significantly suppressed from $\ket{\cal G}_2$.

Moreover, we have calculated the electric transition dipole moment between $\ket{\cal G}_2$ and all states in the first band. The dipole is zero for transitions into all but one of the first band states, and only 18\% of the dipole of a single uncoupled chromophore for the remaining state. By contrast, if the $\ket{D}$ states are not included, the equivalent transition dipole is 41\% of the single chromophore value. Thus, we can conclude that the lowest state of the second band is more protected from optical re-emission if the $\ket{D}$ states are included in the model. This means that including the possibility of doubly-excited sites surprisingly lends a certain degree of extra `dark-state protection' character to the second excitation band, allowing more time for exciton harvesting, and thus improving the efficiency of the device.

There are materials that do not quickly decay through non-radiative transition to the lowest energy state of a given band - these materials then do not obey {\it Kasha's rule}~\cite{kasha50}, which states that optical emission is always from the lowest energy state of a band. For these materials, the reasoning above regarding the composition of the lowest second band state is less relevant. However, in that case, the second step involved in EEA - the non-radiative decay of the $\ket{D}$ state - is also inhibited, and this too leads to an inhibition of EEA and so to more efficient energy harvesting devices.
Indeed, finding materials that violate Kasha's rule for novel optoelectronic applications such as this is a very active area of research~\cite{yanagi10, ghosh06}. 

As discussed above, we expect the $\ket{\cal G}_2$ to have suppressed EEA, but we do not expect it to disappear completely. We therefore look at its impact by introducing EEA it into our model by means of one-way Lindblad operators taking population from all doubly or higher excited states to the highest energy (ladder) state in the single excitation subspace. In Fig.~\ref{fig:annihilation} we show power-voltage characteristics with such EEA taken into account. Each plot compares the ratchets, FD and NP models, for the value of the annihilation rate indicated above it.

As expected, neither the NP case nor the FD scenario are visibly affected. As annihilation increases, the maximal current achieved by the ratchets case eventually reduces to that of FD --- but, importantly, the ratcheting advantage only fully vanishes once the exciton-exciton annihilation rate exceeds the spontaneous emission rate by an order of magnitude or more. On close numerical inspection, FD and NP also suffer very slightly under EEA, because in both those cases the ladder of bright states remains available and there is thus slight chance of multiple excitation. However, unlike in the case of non-radiative recombination, annihilation never kills the dark-state protection advantage.

\section{Demonstration Systems}

In this section we mention some possible systems in which the action of ratchet states might be initially detected.

Ideally, we require a system composed of at least four coupled, radiation-absorbing elements, and where the spacing between each element is either much less than the wavelength of light, or else where the elements can be placed such that the radiation with which they interact is in phase with all of them. This latter condition allows for the possibility of using systems placed an integer number of wavelengths apart, and would include, for example, superconducting qubits interacting with microwave cavity fields. For example, two such qubits have been used to demonstrate super-radiance~\cite{wallraff14}. Including two more would enable ratchets states to be reached, initially by driving the bright excitation and then by altering single qubit phases through local manipulation.

A system closer in spirit to what we have described here would involve coupled optical absorbers. For example,  NV$^-$ centers in diamond are established as systems on which to test quantum information protocols~\cite{waldherr14} and fundamental features of quantum theory~\cite{hensen15}.  NVs can be individually addressed optically, and have spin selective coupling to light that allows optical measurement of their spin state. Moreover, they have been produced in patterns with a precision of around $\lambda/10$, where $\lambda$ is the wavelength at which they emit~\cite{awschalom11}. The prospect of collective photon emission (superrradiance) with multiple NV centres has recently been studied~\cite{hill2016}.

NVs have strain dependent splitting so would have some disorder, though as we have seen in Section~\ref{sec:disorder} above, this does not mean that ratchets states lose all function. However, it may be that silicon vacancies (SiVs) in diamond~\cite{sipahigil16}, or alternatively (di)vacancies in SiC, make better test systems. For example, SiVs in diamond can be placed with $\lambda/20$  precision and are very stable, with inhomogeneous broadening heavily suppressed by their inversion symmetry. For all these kinds of emitter, the F\"orster coupling could be enhanced using a waveguide or cavity to promote the interaction of several centers with a common optical mode~\cite{kurizki96}. Transient luminescence and absorption techniques could be used to track the light excitation in time as it moves through the system, and demonstrate behavior consistent with ratcheting.

Quantum dots are another system that can exhibit strong optical coupling, and could for example be arranged in sub-wavelength patterns using DNA origami~\cite{bui10}. Evidence of collective photon emission of lateral quantum dot ensembles has also already been reported \cite{scheibner2007}.

Of course, the final device likely to be used in a real photocell would be a molecular system. Such systems can be made with high reproducibility, and such that they are very closely spaced with respect to their emission wavelength. H-aggregates of chromophores have a positive F\"orster coupling that shifts the bright (absorbing) state up in energy, as is the case in our model~\cite{mchale12}. Alternatively,
light-harvesting $\pi$-conjugated supramolecular rings can be made with twelve or more repeating units~\cite{anderson11}, specifically to mimic the ring structures found in biology. Indeed, the symmetry of this structure generates a dark $S_0-S_1$ transition in the absence of vibrational distortion~\cite{sprafke11}. Further design optimization of both the individual systems and rigidity of the ring could result in a system ready to meet the requirements for ratcheting.

\end{document}